\documentclass[12pt]{article}

\usepackage{eucal}
\usepackage{color}

\usepackage{tikz}
\usepackage{amsmath,bbm}
\usepackage{amssymb}
\usepackage{epic}

\usepackage{graphicx}
\usepackage{pict2e}

   \usepackage{pdfsync}

   \usepackage{lipsum}

\title{Probability distributions related to tilings of non-convex Polygons}

\author{Mark Adler\thanks{2000
{\em Mathematics Subject Classification}. Primary:
60G60, 60G65, 35Q53; secondary: 60G10, 35Q58. {\em Key
words and Phrases}:Lozenge tilings, non-convex polygons, kernels. \newline
 \hspace*{.5cm}${}^*$Department of Mathematics, Brandeis University,
Waltham, Mass 02453, USA. E-mail: adler@brandeis.edu. The support of a Simons Foundation Grant \# 278931 is gratefully acknowledged. M.A. thanks the Simons Center for Geometry and Physics for its hospitality.
  }
~~~~~ Pierre
van Moerbeke\thanks{ Department of Mathematics,
Universit\'e Catholique de Louvain, 1348 Louvain-la-Neuve, Belgium
and Brandeis University, Waltham, Mass 02453, USA. E-mail: pierre.vanmoerbeke@uclouvain.be . The support of a Simons Foundation Grant 
  \# 280945 is
 gratefully acknowledged. PvM thanks the Simons Center for Geometry and Physics, Stony Brook, and the Kavli Institute of Physics, Santa Barbara, for their hospitality.
\hspace*{1cm} ~~\newline  Published in: Journal of Math. Phys. {\bf 59}, 091418, 21 pp.(2018)
 }
}

\date{}

\newcommand{\MAT}[1]{\left(\begin{array}{*#1c}}
\newcommand{\mat}{\end{array}\right)}
\newcommand{\qed}{\leavevmode\unskip\nobreak\penalty200\hskip2pt\null
\nobreak\hfill\rule{1.1ex}{1.1ex}
\medbreak }

\newcommand{\x}{ \textcolor[rgb]{0.00,0.00,1.00}{\Large \! /\!\!  /\!\! /\!\!   /\!\! /\!\! /\! }
}

\newcommand{\I}{{\rm i}}
\newcommand{\AR}{{\cal A}}
\newcommand{\CR}{{\cal C}}

\newcommand{\RR}{{\cal R}}

\newcommand{\BC}{{\mathbb C}}

\newcommand{\BH}{{\mathbb H}}

\newcommand{\BP}{{\mathbb P}}

\newcommand{\BV}{{\mathbb V}}

\newcommand{\BZ}{{\mathbb Z}}

\newcommand{\iy}{\infty}

\newcommand{\al}{\alpha}

\newcommand{\Id}{\mathbbm{1}}

 \newcommand{\bl}{\begin{aligned}}
  \newcommand{\el}{\end{aligned}}

\newenvironment{proof}{\medskip\noindent{\it Proof:\/} }{\qed}

\newcommand{\la}{\langle}
\newcommand{\ra}{\rangle}
\newcommand{\ga}{\gamma}
\newcommand{\Ga}{\Gamma}

\newcommand{\Dt}{\Delta}
 
\newcommand{\sg}{\sigma}

\newcommand{\BR}{{\mathbb R}}

\newcommand{\dis}{\displaystyle}

\def\be#1\ee{\begin{equation}#1\end{equation}}
\def\bea#1\eea{\begin{eqnarray}#1\end{eqnarray}}
\def\bean#1\eean{\begin{eqnarray*}#1\end{eqnarray*}}

 \newtheorem{definition}{Definition}[section]
 \newtheorem{theorem}[definition]{Theorem}
 \newtheorem{lemma}[definition]{Lemma}
 \newtheorem{corollary}[definition]{Corollary}
 \newtheorem{proposition}[definition]{Proposition}

\catcode `!=11

\newdimen\squaresize
\newdimen\thickness
\newdimen\Thickness
\newdimen\ll! \newdimen \uu! \newdimen\dd! \newdimen \rr! \newdimen
\temp!

\def\sq!#1#2#3#4#5{%
\ll!=#1 \uu!=#2 \dd!=#3 \rr!=#4
\setbox0=\hbox{%
 \temp!=\squaresize\advance\temp! by .5\uu!
 \rlap{\kern -.5\ll!
 \vbox{\hrule height \temp! width#1 depth .5\dd!}}%
 \temp!=\squaresize\advance\temp! by -.5\uu!
 \rlap{\raise\temp!
 \vbox{\hrule height #2 width \squaresize}}%
 \rlap{\raise -.5\dd!
 \vbox{\hrule height #3 width \squaresize}}%
 \temp!=\squaresize\advance\temp! by .5\uu!
 \rlap{\kern \squaresize \kern-.5\rr!
 \vbox{\hrule height \temp! width#4 depth .5\dd!}}%
 \rlap{\kern .5\squaresize\raise .5\squaresize
 \vbox to 0pt{\vss\hbox to 0pt{\hss $#5$\hss}\vss}}%
}
 \ht0=0pt \dp0=0pt \box0
}

\def\vsq!#1#2#3#4#5\endvsq!{\vbox to \squaresize{\hrule
width\squaresize height 0pt%
\vss\sq!{#1}{#2}{#3}{#4}{#5}}}

\newdimen \LL! \newdimen \UU! \newdimen \DD! \newdimen \RR!

\def\vvsq!{\futurelet\next\vvvsq!}
\def\vvvsq!{\relax
  \ifx     \next l\LL!=\Thickness \let\continue=\skipnexttoken!
  \else\ifx\next u\UU!=\Thickness \let\continue=\skipnexttoken!
  \else\ifx\next d\DD!=\Thickness \let\continue=\skipnexttoken!
  \else\ifx\next r\RR!=\Thickness \let\continue=\skipnexttoken!
  \else\def\continue{\vsq!\LL!\UU!\DD!\RR!}%
  \fi\fi\fi\fi
  \continue}

\def\skipnexttoken!#1{\vvsq!}

\def\place#1#2#3{\vbox to 0pt{\vss
\rlap{\kern#1\squaresize
  \raise#2\squaresize\hbox{$#3$}}
\vss}}

\catcode `!=12

\newsavebox{\foobox}
\newcommand{\slantbox}[2][.5]
  {%
    \mbox
      {%
        \sbox{\foobox}{#2}%
        \hskip\wd\foobox
        \pdfsave
        \pdfsetmatrix{1 0 #1 1}%
        \llap{\usebox{\foobox}}%
        \pdfrestore
      }%
  }

\begin{document}

\sloppy
\maketitle


 


  
 
 
 


 \begin{abstract} This paper is based on the study of random lozenge tilings of non-convex polygonal regions with interacting non-convexities (cuts) and the corresponding asymptotic kernel  as in [3] and [4] (discrete tacnode kernel). Here this kernel is used to the find the probability distributions and joint probability distributions for the fluctuation of tiles along lines in between the cuts. These distributions are new.
  \end{abstract}
  
  
  \newpage

   \vspace*{-9cm}
  
  \par\vspace*{.35\textheight}{\centering Dedicated to the memory of Ludvig Faddeev  \par}

 
    
   \section{The discrete tacnode kernel and main result}

   Domino or lozenge tilings of large geometric shapes constitute a rich source of new statistical phenomena:  
   they have sufficient complexity to have interesting features and yet are simple enough to be tractable! Most models studied sofar display two phases, a solid phase with a bricklike pattern, and a liquid phase, for which the correlations decay polynomially with distance. More recently, new models were considered having an additional phase, a gas phase, for which the correlations decay exponentially with distance. This paper will deal with a model having the two phases, solid and liquid. 
   
   A celebrated example goes back to MacMahon \cite{McMa} in 1911, who found a simple combinatorial formula for the number of lozenge tilings of a hexagon of sides $a,b,c,a,b,c$. This model has been widely studied and extended from the macroscopic point of view, but also from the microscopic point of view \cite{Cohn,Jo02b,Jo05b,Ciucu}. For tilings of hexagons, everything is known:  when the size gets large, an {\em arctic ellipse}, inscribed in the hexagon, separates the liquid phase and the solid phases appearing in the six corners of the hexagon. The liquid phase behaves like a Gaussian Free Field, the statistical fluctuations of the tiles along the ellipse fluctuate according to the Airy process\cite{Jo05c}. The tiles in the neighborhood of the tangency points of the arctic ellipse with the hexagon behave as the eigenvalues of the consecutive principal minors of a GUE-matrix ({\em GUE-minor process}) \cite{JN}. They are all {\em universal distributions}, in the sense that they have been found in entirely different circumstances as well. They are also known to occur at critical points along the boundary between phases. The universal distributions are "{\em integrable}": many of them relate to known integrable systems, like KdV equation, the Boussinesq equation, Toda lattices, etc... or they can be treated by means of Riemann-Hilbert methods. 
   
 {\em   This paper written in memory of Ludvig Faddeev is a tribute to his pioneering contributions to the integrable field. In 1970-71, he gave a lecture at Rockefeller University in NY on the KdV equation, showing that KdV is a completely integrable Hamiltonian system, and that the map to the spectral data is symplectic. His brilliant lecture triggered the interest and inspiration of one of the authors of this paper (PvM): thank you, Ludvig! }
   
   Domino tilings of Aztec diamonds have also been extensively studied from the combinatorial point of view and from the microscopic point of view  \cite{EKLP,EKLP2,JPS,Pro:03,Jo02b,Jo05c}.
   
   \newpage
  
     \vspace*{-3.0cm}
 \setlength{\unitlength}{0.015in}\begin{picture}(0,0)

   {\makebox(450,-480) {\rotatebox{0
    } {\includegraphics[width=240mm,height=270mm]
    {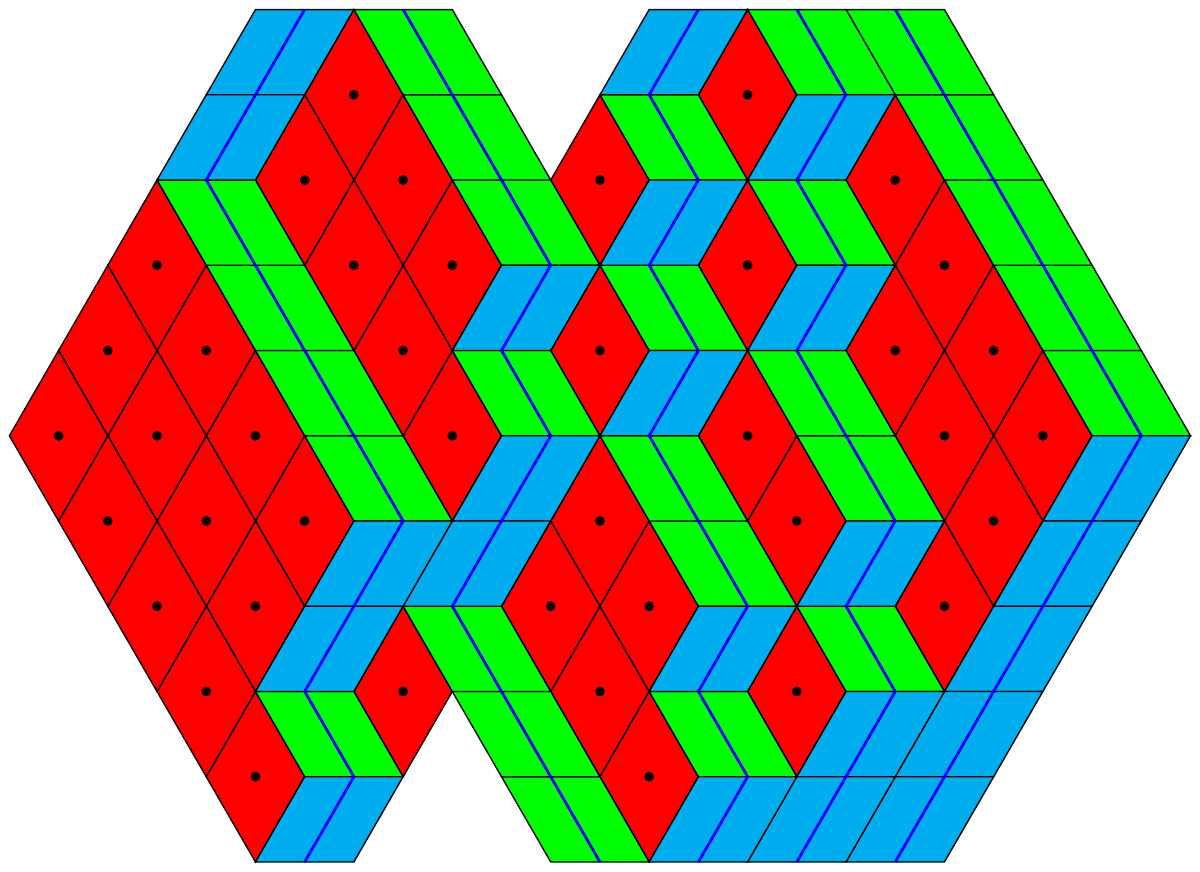} }}}
    
  \put(-300,-225){\makebox(0,0) {\footnotesize (Courtesy of Antoine Doeraene)}}
        
    \multiput(450,-178)(4,0){27}{$\mbox{\tiny$.$}$}
    
     \multiput(-632,159)(4,0){27}{$\mbox{\tiny$.$}$}

    \end{picture}
    
     \vspace*{9cm}
 Fig.1: Lozenge tiling of a hexagon with cuts.
   

\vspace*{-1cm}

  \setlength{\unitlength}{0.017in}\begin{picture}(0,60)
\put(145,-70){\makebox(0,0) {\rotatebox{-90}{\includegraphics[width=100mm,height=120mm]
 {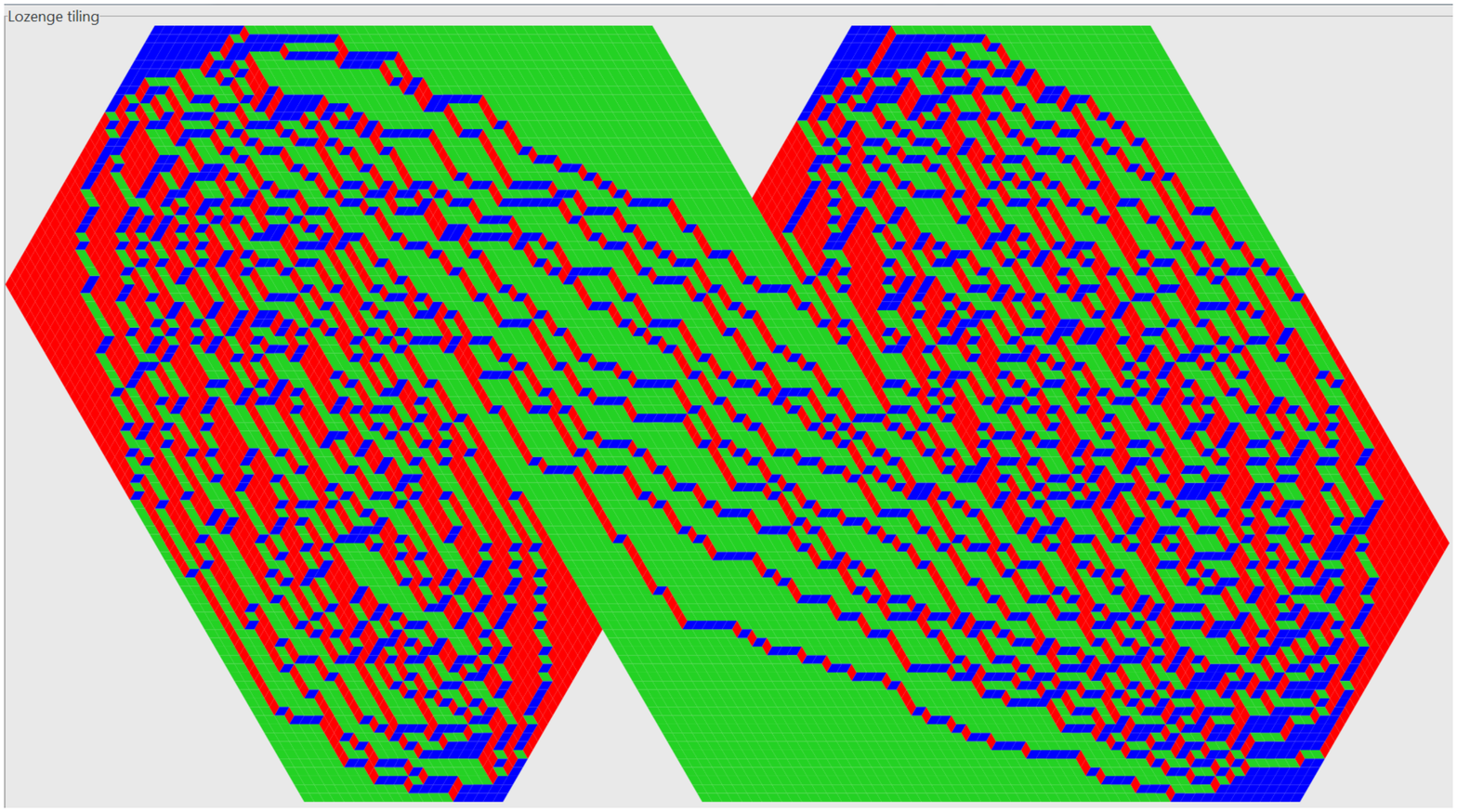}}}}
 
   \put(145,-170){ \line(-0.503, 1){100}}
      
        \put(219,-170){ \line(-0.503, 1){100}}
        
         \put(141,-160){ \vector(1, 0){73}}
          \put(141,-160){ \vector(-1, 0){1}}
         \put(176,-168){\makebox(0,0){ $\rho$} }


 
  

 
 
  \end{picture}
  
  \vspace*{8cm}
 Fig.2: Computer simulation of lozenge tilings of a hexagon, with $ b = 30,~ c = 60,~n_1 = 50, n_2 = 30, ~d = 20,~m_1=20,~m_2=60$. The strip $\{\rho\}$ of width $\rho=n_1-m_1+b-d= 10$ contains $r=b-d$ paths of blue and red tiles.

\newpage

    \newpage
    
    \vspace*{-4cm}  
    
  \hspace*{-4cm}
  \slantbox[-.60]{    \setlength{\unitlength}{0.015in}\begin{picture}(0,0)

   {\makebox(450,-480) {\rotatebox{0
    } {\includegraphics[width=200mm,height=270mm]
    {tiling_Antoine.pdf} }}}
     \end{picture}  }
     
     \vspace*{8cm}
      
      Fig. 3: Affine transformation of Fig.1. 

   
    \vspace*{-5cm}

\setlength{\unitlength}{0.017in}\begin{picture}(0,170)
\put(150,-70){\makebox(0,0) {\rotatebox{0}{\includegraphics[width=135mm,height=200mm]
 {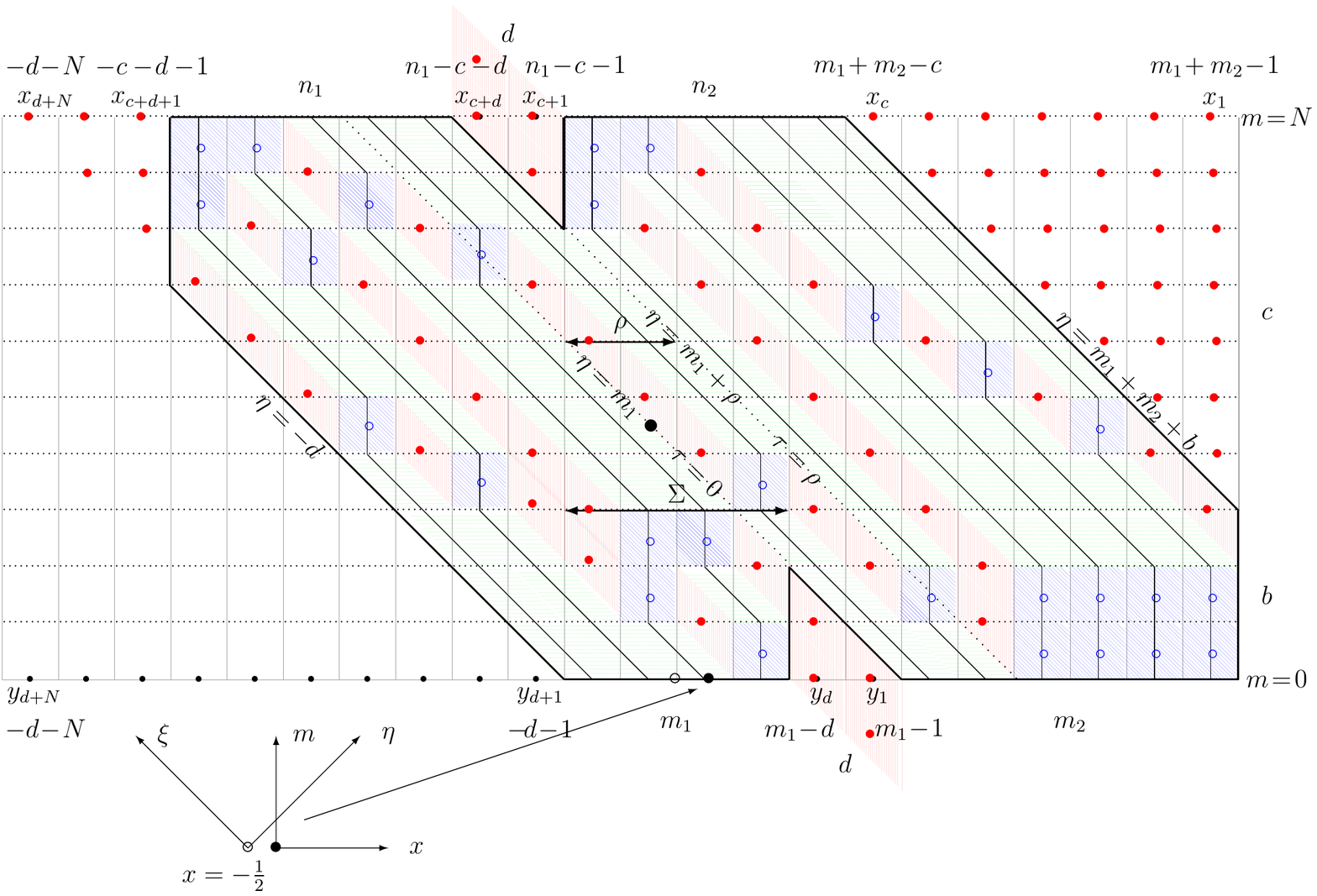} }}}
     
\end{picture}

\vspace*{8.8cm}

Fig. 4. Tiling of a hexagon with two opposite cuts of equal size ({\em Two-cut case}), with red, blue and green tiles. Here $d=2$, $n_1=n_2=5,~m_1=4,~m_2=6,~ b=3,~c=7$, and thus $~ r=1,~ \rho=2.$ The $(m,x )$-coordinates have their origin at the black dot on the bottom-axis $m=0$ and the $(\eta,\xi )$-coordinates at the circle given by $(m,x )=(0,-\tfrac 12 )$. 
%

\newpage
  

   Do lozenge tilings of different geometric shapes lead to other universal distributions? Yes, they do. This is what this paper is about. Indeed, consider a lozenge tiling of a hexagon with cuts along opposite sides, with blue, red and green tiles as pictured in Fig. 1. Letting the shape become large and the cuts as well (Fig.2), a computer random simulation shows -roughly speaking- the appearance of two inscribed ellipses in two ``near-hexagons'' connected with paths of blue and red tiles in a sea of green tiles. These paths traverse a strip obtained by extending linearly two parallel sides of the cuts. The purpose of this paper is to determine the distribution of the blue tiles along lines parallel to the strip and joint distributions along two such lines, for an appropriate scaling of the polygon and the cuts. 
   Domino tilings of non-convex figures have also been studied before; e.g. in \cite{KO,KOS,AJvM,ACJvM,AvM}. This paper is based and is a follow-up of papers \cite{AJvM1,AJvM2}; see also \cite{BD,DJM,Gorin1,OR,Petrov1}
   
   To find out such a statistics for the lozenge tilings, we first need to determine the correlation kernel for the finite problem, in particular showing that the point process of blue tiles (replaced by blue dots in the middle of the tiles) along the parallel lines, mentioned above, is determinantal, with an explicitly given correlation kernel ${\mathbb K}^{\mbox{\tiny blue}}(\xi_1,\eta_1;\xi_2,\eta_2)$, in the variables $\eta, \xi$ (to be explained later). For convenience of coordinates, we perform an affine transformation of Fig. 1 to obtain Fig. 3. This leads us to consider the basic model, as depicted in Fig. 4, with upper-edge $n_1, n_2$ with a cut of size $d$, with lower-edge $m_1, m_2$ also with a cut of size $d$, the two remaining parallel edges having sizes $b$ and $c$. 
   The oblique lines parallel to the strip, named $\{\rho\}$ of width $\rho$ (given in formula (\ref{r,rho}) below), will be parametrized by the (integer) coordinate $-d\leq\eta\leq m_1+m_2+b$, as seen in Fig. 4, with (integer) running variable $\xi$ along those lines. The precise coordinates $(\eta,\xi)$ will be given in (\ref{Lcoord}). It is easily shown that the number of blue dots along the $\rho+1$ parallel lines within and at the boundary of the strip $\{\rho\}$ is always the same and equals $r:=b-d$. So, 
%
 \be\mbox{fixing} ~\left\{ \begin{aligned}
  \rho&=  n_1-m_1+b-d  =m_2-n_2+b-d 
  =\{\mbox{width of the strip $\{\rho\}$}\} 
\\
 r&=b-d=\#\{\mbox{blue dots on the parallel lines in the strip $\{\rho\}$}\} ,
\end{aligned}\right.
\label{r,rho}\ee
we let the following data go to infinity together, according to the following scaling:
   \newline
    {\em (i) Scaling of geometrical data:} letting the size of the cuts $=d\to \iy$, 
$$\begin{array}{lllll}
b=d+r&& c=\kappa d
\\
m_1=\tfrac{\kappa+1}{\kappa-1}
 ( d+\sqrt{\frac{\kappa}{\kappa-1}}  \beta_1\sqrt{d}+ \ga_1) && n_1 = m_1+(\rho-r)   
\\  
m_2=\tfrac{\kappa+1}{\kappa-1}
 ( d+\sqrt{\frac{\kappa}{\kappa-1}}  \beta_2\sqrt{d}+ \ga_2) &&  n_2 = m_2-(\rho-r)  ,  
\\ \\
\beta=-\beta_1-\beta_2,~~&&~ 1<\kappa<3 ,~\beta_i,~\gamma_i=\mbox{free parameters}
 \end{array} $$
%
%
\newline
{\em (ii) Scaling of running variables:}   $(\eta_i,\xi_i)\in \BZ^2\to (\tau_i,\theta_i)\in (\BZ \times\BR ) ,$ about the point  $(\eta_0,\xi_0) $  given by the black dot in Fig. 4, (halfway point along the left boundary of the strip $\{\rho\}$ shifted by $(-\tfrac 12, \tfrac 12)$), upon setting $a=2\sqrt{\frac{\kappa}{\kappa-1}}$,
\be\begin{aligned}
 (\eta_i,\xi_i) &=(\eta_0,\xi_0)+(\tau_i,~\tfrac {\kappa+1}a (\theta_i + \beta_2)\sqrt{d} )\mbox{    with   }
  (\eta_0,\xi_0)  =(m_1,N-m_1 -1).
   \end{aligned}
\label{eta-xi}\ee
The main statement of the paper is based on the following asymptotic result:

\begin{proposition} \label{Th1}(Adler, Johansson, van Moerbeke \cite{AJvM1,AJvM2})Given the scaling above, applied to the correlation kernel ${\mathbb K}^{\mbox{\tiny blue}}$ of blue dots, it is shown that 
\be\begin{aligned}
 \lim_{d\to \infty} (-1)^{\tfrac 12 
   (\eta_1+\xi_1-\eta_2-\xi_2)}
&\left(\sqrt{d}\frac{\kappa + 1}{2a}\right)^{\eta_2-\eta_1 
 }
{\mathbb K}^{\mbox{\tiny blue}}( \eta_1 ,\xi_1;\eta_2,\xi_2)\frac 12 \Dt\xi_2
\\
&={\mathbb L}^{\mbox{\tiny dTac}} (\tau_1, \theta_1;\tau_2, \theta_2)d\theta_2,\end{aligned}
 \label{limit}\ee 
where  \be \label{Final0}\begin{aligned}
 {\mathbb L}^{\mbox{\tiny dTac}}  (&\tau_1, \theta_1 ;\tau_2, \theta_2)  
  :=   {\mathbb K}^{\mbox{\tiny GUE}}  ( \tau_1\!-\!\rho, -\theta_1 ;\tau_2\!-\!\rho, -\theta_2)  
\\& +\oint_{\Ga_0}\frac{du} {(2\pi\I)^2}\oint_{\uparrow L_{0+}}  \frac{ dv}{v-u}\frac{v^{\tau_2-\rho }}{u^{\tau_1-\rho }}
\frac{e^{-u^2  -  \theta_1    u  }}
{e^{-v^2 - \theta_2   v  }}
      \frac{ \Theta_r( u, v )-\Theta_r(0,0)} { \Theta_r(0,0)}  
\\& +r\oint_{\uparrow L_{0+}  }    \frac{du} {(2\pi\I)^2} \oint_{\uparrow L_{0+}}dv  \frac{v^{\tau_2-\rho}}{u^{\tau_1}}
\frac{e^{ u^2 -    (\theta_1- \beta     )u  }}
{e^{-v^2 - \theta_2    v  }}
    \frac{ \Theta^+_{r-1}(  u, v )} {  \Theta_r(0,0)} 
    \\& +\oint_{\Ga_0}\frac{du} {(2\pi\I)^2}\oint_{\uparrow L_{0+}}\frac{ dv}{v-u}  \frac{v^{  -\tau_ 1}}{u^{-  \tau_2}}
\frac{e^{ -u^2 +  (\theta_2- \beta  )u  }}{e^{- v^2  + (   \theta_1- \beta    )v  }}
     \frac{ \Theta_r( u  , v )} { \Theta_r(0,0)} 
 \\
 & -\tfrac1{r+1}\oint_{\Ga_0}\frac{du} {(2\pi\I)^2} \oint_{\Ga_0}dv
 \frac{v^{  \tau_2}}{u^{\tau_1- \rho }}
\frac{e^{-u^2 - \theta_1 u  }}
{e^{ v^2 - (\theta_2- \beta    )v  }}
    \frac{ \Theta^-_{r+1}( u, v  )} {  \Theta_r(0,0)}
    \\& =:
  {\mathbb K}^{\mbox{\tiny GUE}}  ( \tau_1\!-\!\rho, -\theta_1 ;\tau_2\!-\!\rho, -\theta_2)   +    \sum_1^4 {\mathbb L}^{\mbox{\tiny dTac}}_i (\tau_1, \theta_1 ;\tau_2, \theta_2) ,
\end{aligned}\ee
 where $ {\mathbb K}^{\mbox{\tiny GUE}}$ is the GUE-minor kernel, and where the $\Theta^{\pm}_k$ are $k$-fold multiple integrals, to be give in (\ref{Theta}). The integrations are taken along  upwards oriented vertical lines $\uparrow L _{0+}$ to the right of a (counterclock) contour $\Gamma_0$ about the origin and with the basic integers $r$ and $\rho$. 
 Finally the kernel satisfies the following involution:
\be \label{invol}
{\mathbb L}^{\mbox{\tiny dTac}} (\tau_1, \theta_1;\tau_2, \theta_2)
=
{\mathbb L}^{\mbox{\tiny dTac}} (\rho-\tau_2, \beta-\theta_2;\rho-\tau_1, \beta-\theta_1).
\ee
   
  \end{proposition}


Given $\tau\in \BZ_{\geq 0}$, define:
\be{\bf z}^{(\tau)}
=( z_n ^{(\tau)}\leq \dots\leq z_1 ^{(\tau)}),\mbox{   where  } n=n_\tau:=(\tau-\rho)_{>0}+r
 \label{zn}\ee
together with the polytope (truncated cone) of interlacing sets\footnote{ with
$$\bl
&  z \prec u \mbox{ for $z,u \in \BR^n$ meaning  } 
 z_n\leq  u_n\leq z_{n-1} \leq u_{n-1} \leq \dots\leq z_1 \leq  u_1 ,
\\
&   z \prec u \mbox{ for $z  \in \BR^n$ and $u\in \BR^{n+1}$ meaning  } 
 u_{n+1}\leq z_n\leq  u_n\leq z_{n-1} \leq u_{n-1} \leq \dots\leq z_1 \leq  u_1. \el $$
 } for given ${\bf x}= {\bf x}^{(\tau_1 )}$ and $ {\bf y}={\bf y}^{(\tau_2 )}$
 $$
 \CR(\tau_1,{\bf x};\tau_2,{\bf y}):=\left\{\begin{array}{llllll}
 {\bf z}^{(\tau_1 )}\prec {\bf z}^{(\tau_1+1)}\prec\dots\prec {\bf z}^{(\tau_2-1)}\prec {\bf z}^{(\tau_2 )}, 
 \\ \mbox{with  } 
 {\bf z}^{(\tau_1 )}={\bf x}\mbox{   and  }{\bf z}^{(\tau_2 )}={\bf y} 
 \end{array}
 \right\}
 $$
with uniform measure on $ \CR(\tau_1,{\bf x};\tau_2,{\bf y})$ (Lebesgue measure)
$$d\mu_{{\bf x}\bf y}( {\bf z}^{(\tau_1+1 )}, \dots,  {\bf z}^{(\tau_2 -1)})
=\left(\prod_{\tau_1< \tau< \tau_2}
 d{\bf z} ^{(\tau)}\right)
 \Id_{ {\bf z}^{(\tau_1 )}\prec \dots\prec  {\bf z}^{(\tau_2 )}}
\mbox{   with  } d{\bf z}^{(\tau)}=\prod_{i=1}^{n_\tau} dz_i^{(\tau)}.$$
The volume of $\CR(\tau_1,{\bf x};\tau_2,{\bf y})$ is then given by 
\be\label{vol}
 \mbox{Vol}(\CR(\tau_1,{\bf x};\tau_2,{\bf y}))=\int_{\CR(\tau_1,{\bf x};\tau_2,{\bf y})}  d\mu_{{\bf x}\bf y}( {\bf z}^{(\tau_1+1 )}, \dots,  {\bf z}^{(\tau_2 -1)}).
 \ee
%
In order to state the main theorem, we define, besides the usual Wronskian $\Dt_n(y)  $, Wronskian-like determinants, for $x:=(x_1,\dots,x_n)$. The definition will depend on whether $0\leq \tau\leq \rho$ or $\tau>\rho$. When $0\leq \tau\leq \rho$, we use the first expression $\widetilde \Dt_{r,\tau}^{(\tau\leq \rho)}(x )$ and when $\tau>\rho$, the matrix in $ \widetilde \Dt_{n,\tau}^{(\tau>\rho )  }$ has a regular Wronskian part of size equal to the distance between $\tau$ and the strip $\{\rho\}$, where one adjoins the matrix in $\widetilde \Dt_{r,\tau}^{(\tau\leq \rho)}$ accounting for the strip:
\be  \label{Deltatilde}
\bl
\widetilde \Dt_{r,\tau}^{(\tau\leq \rho)}(x )=\det\left(\begin{array}{ccc}
 \Phi_{\tau-1 }(x_j )
 \\
 \vdots
 \\
 \Phi_{\tau -r }(x_j )
\end{array}\right)_{j=1,\dots,r},~~
 \widetilde \Dt_{n,\tau}^{(\tau>\rho )  } (x )=\det
 \left(\begin{array}{ccc}
 1 
 \\x_j
 \\ \vdots
 \\
 x_j^{\tau-\rho-1}
 \\
 \Phi_{\tau-1 }(x_j )
 \\
 \vdots
 \\
 \Phi_{\tau -r }(x_j )
 \end{array}
 \right)_{j=1,\dots,n}
\hspace{-2cm}\begin{array}{l} 
 \\ \\ \\ \\ \\  \end{array} 
 \hspace*{1cm},\el\ee 
where $\Phi_n (\eta)$ is a Gaussian-type integral along a vertical complex line given by
\be \label{Phi}\Phi_n (\eta)  :=\frac{1}{2\pi \I} \int_L \frac{e^{v^2+2\eta v}}{v^{n+1}} dv. \ee
Also define\footnote{In the formula below, we set ${\bf x}+\tfrac \beta 2:=(x_{n_1}+\tfrac \beta 2,\dots,x_{ 1}+\tfrac \beta 2)$.} 
\be
D(\tau_1,{\bf x}^{(\tau_1)};\tau_2,{\bf y}^{(\tau_2)}):=
\left\{ \begin{array}{ll}
C_{\tau_1,\tau_2,r}\widetilde \Dt_{n_1,\tau_1}^{(\tau_1>\rho )  } ({\bf x}+\tfrac \beta 2 ) 
\left(\prod_{i=1}^{n_2} \frac{e^{- y_i^2  }}{ \sqrt{ \pi} }\right) \Dt_{n_2}({\bf y})
\\ \\
\hspace*{2cm}\mbox{  for $\rho\leq \tau_1\leq \tau_2$}
\\ \\ \\
 C'_{\tau_1,\tau_2,r}
 \widetilde\Dt_{n_1,\tau_1}^{(\tau_1\leq\rho)}({\bf x}+\tfrac \beta 2)
 \widetilde\Dt_{n_2,\rho-\tau_2}^{(\tau_2\leq\rho)}(-{\bf y} )
 \\ \\
\hspace*{2cm}\mbox{  for $0\leq \tau_1\leq \tau_2\leq \rho$ and $n_1=n_2=r$,}
\end{array}\right.\label{D}\ee  
with 
\be\bl
C_{\tau_1,\tau_2,r}&:=\frac{(-1)^{\frac 12 n_1(n_1-1)}2^{-n_1r}}{(-1)^{\frac 12 n_2(n_2-1)}2^{-n_2r}}
\frac{\sqrt{2}^{(n_2-r)(n_2-r-1)}}{\prod_{k=0}^{n_1-r-1}k!}
\det(\widetilde \Ga^{(r-1)}_{k,\ell})_{0\leq k,\ell\leq r-1} 
\\
C'_{\tau_1,\tau_2,r}&:=
   2^{r(\tau_2-\tau_1+1)} \det\left(   \widetilde \Ga^{(r-1)}_{k,\ell}\right)_{0\leq k,\ell \leq r-1}, \el \label{C'}\ee
where $\widetilde \Ga^{(r-1)}_{k,\ell}$ are $r-1$-fold integrals given later in (\ref{Gamma}).
 Throughout the paper, it will be more convenient to replace the variables $\theta_i$ in the kernel ${\mathbb L}^{\mbox{\tiny dTac}} (\tau_1, \theta_1;\tau_2, \theta_2)$ by new variables $x$ and $y$ already used in (\ref{Deltatilde}), (\ref{Phi}) and (\ref{D}), namely : \be
 \theta_1=-2x\mbox{   and   }  \theta_2=-2y.
 \label{theta-xy}
 \ee
  The main statement reads as follows: 

 \begin{theorem} \label{Th1.2} The distribution of the blue tiles at position ${\bf x}={\bf x}^{(\tau )}  $ along level $\tau\geq 0$ and their joint distribution at ${\bf x}={\bf x}^{(\tau_1 )}  $ and ${\bf y}={\bf y}^{(\tau_2 )}  $ along levels $0\leq \tau_1\leq \tau_2$ are given by:
\be \label{Prob'}\bl
\BP&\left(\begin{array}{l} {\bf x}^{(\tau )}\in d{\bf x}^{ }   \end{array}\right)
 =D(\tau ,{\bf x};\tau ,{\bf x})
d{\bf x},~~~~  \mbox{     for   } \tau\geq 0
\\ \\
\BP&\left(\begin{array}{l} {\bf x}^{(\tau_1)}\in d{\bf x}   \mbox{   and  }  {\bf y}^{(\tau_2)}\in d{\bf y} \end{array}\right)
 =D(\tau_1,{\bf x};\tau_2,{\bf y})
\mbox{Vol}(\CR(\tau_1,{\bf x};\tau_2,{\bf y}))
d{\bf x} 
d{\bf y},
\\  &\hspace*{4cm}~~~~~ \mbox{     for  $0\leq \tau_1<\tau_2\leq \rho$~~~ or~~~ $\rho<\tau_1<\tau_2 $.}
\el\ee
The involution (\ref{invol}) enables one to obtain similar formulas for $\tau_2<\tau_1<0$.
\end{theorem}

For future use, the density of the probabilities above will be denoted by $p\left(\tau,{\bf x}\right)$ and $p\left(\tau_1,{\bf x}_1;\tau_2,{\bf x}_2\right)$.

\begin{corollary} \label{Cor1.3} The joint probability of the blue tiles at all levels between $\tau_1$ and $\tau_2$ for $\tau_1<\tau_2$ equals
\be\label{Gibbs}\bl
  \BP&\left(  {\bf x}^{(\tau_1)}\in d{\bf x}^{(\tau_1)} ,~
  {\bf z}^{(\tau_1+1)}\in d{\bf z}^{(\tau_1+1)} , 
\dots, {\bf z}^{(\tau_2-1)}\in d{\bf z}^{(\tau_2-1)} ,~{\bf y}^{(\tau_2)}\in d{\bf y}^{(\tau_2)} \right)
\\&=D(\tau_1,{\bf x}^{(\tau_1)};\tau_2,{\bf y}^{(\tau_2)}) d{\bf x}^{(\tau_1)}~
d{\bf x}^{(\tau_2)}
d\mu_{{{\bf x}^{(\tau_1)}}{\bf y}^{(\tau_2)}}( {\bf z}^{(\tau_1+1 )}, \dots,  {\bf z}^{(\tau_2 -1)}). \el \ee
 
 \end{corollary}

This is a consequence of the {\em Gibbs property} which expresses the fact that given the positions of the blue tiles along two levels $\tau_1<\tau_2$, the positions of the blue tiles in between are uniformly distributed, while respecting the interlacing. This is easily seen to hold in the discrete case of Fig. 4, using the height function given in \cite{AJvM1} and similar arguments for the Gibbs property given in \cite{ACJvM}, section 4. The property is obviously maintained in the scaling limit of Proposition \ref{Th1}.

\bigbreak

 \noindent{\em \underline{Remark 1}:}~The joint probability is computed here for the two cases $0\leq \tau_1<\tau_2\leq \rho$ and $\rho<\tau_1<\tau_2 $. It is still an open problem to obtain the joint probability for other circumstances, like e.g., $0\leq \tau_1<\rho<\tau_2$ or $\tau_1<0<\tau_2$.
 
 \bigbreak

\noindent{\em \underline{Remark 2}:}~When $\rho=r$, i.e., when $n_1=m_1$ and $n_2=m_2$ from (\ref{r,rho}), the limiting distribution (\ref{Prob'}) has appeared in domino tilings of overlapping Aztec diamonds \cite{AvM}. It suggests that the discrete tacnode kernel (\ref{Final0}) is more general than the tacnode kernel for overlapping Aztec diamonds obtained in \cite{ACJvM}.

     \newpage

\section{Some background on the discrete tacnode kernel}

  The affine map of Fig.1 leading to Figs. 3 or 4  will enable us to describe the tilings by convenient systems of coordinates $(m,x)$ and $(\eta, \xi)$, related by
    \be\eta=m+x+\tfrac 12,~~~\xi=m-x-\tfrac 12 ~~~~~ \Leftrightarrow ~~~m =\tfrac 12(\eta+\xi),~~x=\tfrac 12 (\eta-\xi-1) .\label{Lcoord}\ee
    where $1\leq m\leq N$ are the horizontal (dotted) lines in Fig. 4 and $x$ the running integer variable along those lines. The integers along the bottom line $\notin {\bf P}$ are labeled by $y_1>y_2>\dots>y_{d+N}$, with $y_1=m_1-1,~y_d=m_1-d,~y_{d+1}=-d-1,~y_{d+N}=-d-N$. The integers along the top line $\notin {\bf P}$ are labeled by $x_1>x_2>\dots>x_{d+N}$, with $x_1=m_1+m_2-1, ~x_c=m_1+m_2-c, ~x_{c+1}=n_1-c-1,~x_{c+d}=n_1-c-d,~x_{d+N}=-d-N$.
   As mentioned, the coordinate $-d\leq\eta\leq m_1+m_2+b$ parametrizes the oblique lines parallel to the strip $\{\rho\}$, with $\xi$ being the running variable along those lines. Notice that the cuts are taken care of by filling them up with red tiles, as seen in Fig. 4. 
    
 In \cite{AJvM1}, we considered two different discrete-time processes, which we show to be determinantal : a ${\mathbb K}^{\mbox{\tiny red}}$-point process of red tiles situated along the horizontal lines $0\leq m\leq N$,  and the aforementioned ${\mathbb K}^{\mbox{\tiny blue}}$-point process of blue tiles along the oblique lines $-d+1\leq \eta\leq m_1+m_2+b-1$; see Fig. 4.  The ${\mathbb K}^{\mbox{\tiny blue}}$-kernel is the determinantal processes of blue dots belonging to the intersection of the parallel oblique lines $x+m=k-\tfrac 12$  with the horizontal lines $m=\ell-\tfrac 12$ for $k,\ell \in \BZ$; so
the blue dots are parametrized by $(\eta,\xi  )=(k,2\ell-k-1 )\in \BZ^2$, with $( k,\ell)$   as above. It follows that the $(\eta,\xi )$-coordinates of the blue dots satisfy $\xi+\eta=1,3,\dots,2N-1$. In \cite{AJvM1,AJvM2} it was shown that the two kernels ${\mathbb K}^{\mbox{\tiny red}}$ and ${\mathbb K}^{\mbox{\tiny blue}}$ are intimately related through the fact that ${\mathbb K}^{\mbox{\tiny red}}$ is, up to a sign, the inverse of the Kasteleyn matrix \cite{Kast} (adjacency matrix) for the dimer model constructed on the honeycomb lattice. One first computes the ${\mathbb K}^{\mbox{\tiny red}}$-kernel and then deduce the ${\mathbb K}^{\mbox{\tiny blue}}$-kernel from the Kasteleyn matrix. 

As pointed out in Theorem \ref{Th1}, the discrete tacnode kernel $ {\mathbb L}^{\mbox{\tiny dTac}}$-kernel (\ref{Final0}) is the scaling limit of the $ {\mathbb K}^{\mbox{\tiny blue}}$-kernel and is given in terms of the GUE-minor kernel, and multiple integrals $\Theta_k$, given by:
%
 \be   \begin{aligned}
     {\mathbb K}^{\mbox{\tiny GUE}}  ( n_1, x_1 ;n_2, x_2)&:= - 
{\mathbb H}^{n_1-n_2}(  x_1-  x_2) 
+\oint_{\Ga_0}\frac{du} {(2\pi\I)^2}\oint_{\uparrow L_{0+}}  \frac{ dv}{v-u}\frac{v^{n_2}}{u^{n_1}}
\frac{e^{-u^2  + x_1    u  }}
{e^{-v^2 +x_2  v  }}
\\
 \Theta_r( u, v)&:=\left[     
\prod_1^r\oint_{\uparrow L_{0+}}\frac{e^{    2w_\al^2+\beta   w_\al}}{    w_\al  ^{\rho }}
 ~\left(\frac{v\!-\!w_\al}{u\!-\!w_\al}\right) \frac{dw_\al}{2\pi \I}\right]\Dt_r^2(w_1,\dots,w_r)
%
 \\  \Theta^{\pm}_{r\mp1}( u, v)     &:=\left[     
\prod_1^{r\mp 1}\oint_{\uparrow L_{0+}}\frac{e^{    2w_\al^2+\beta  w_\al}}{    w_\al  ^{\rho }}
 ~\left( ({v\!-\!w_\al} )\ ({u\!-\!w_\al})\right)^{\pm 1} \frac{dw_\al}{2\pi \I}\right]
 \\
 &~~ ~~\qquad~ ~\qquad \qquad~~~~~~~~~~~~~~~~~~~~~\Dt_{r\mp1}^2(w_1,\dots,w_{r\mp1}).
 \\
 \BH^{m}(z)&:=\frac{z^{m-1}}{(m-1)!}\Id _{z\geq 0}\Id_{m\geq 1},~~~~(\mbox{Heaviside function})
 \label{Theta}
\end{aligned}\ee

 In Section 2 of  \cite{AJvM2}, it was shown that in the scaling limit, each level $\tau\geq 0$ carries\footnote{with $(\tau-\rho)_{>0}:=\Id_{\tau>\rho}(\tau-\rho)$.} \be n=(\tau-\rho)_{>0}+r\label{n}\ee blue dots. 
  That is to say within the strip and on its boundary the oblique lines $0\leq \tau\leq \rho$ carry $r$ blue dots and then on either side, the numbers go up by $1$. 

 \section{Volume of truncated polytopes}

 Define the standard Hermite polynomials $H_n$, related polynomials $\tilde H_n$ and $P_n$ :
\be
\begin{aligned}
H_n(x)&:= n! \oint_{\Gamma_0}e^{-z^2+2xz}\frac {dz}{2\pi \I z^{n+1}}= 2^{n+1}\sqrt{\pi}e^{x^2} \int_{L}e^{w^2-2xw}w^n\frac{dw}{2\pi \I}
=(2x)^n+\ldots\\
\widetilde H_n(x)&:=\frac {H_n(x)}{n!}=\tfrac{2^n}{n!}x^n+\dots ,~~ 
\widehat H_n(x) :=\frac {H_n(x)}{2^{n+1}} =\tfrac12 x^n+\dots,~~ 
\bar H_n(x):=\frac{H_n(x)}{2^n n!}=\tfrac{x^n}{n!} +\dots
\\  
%
P_n(x)&:=  \begin{aligned}&\frac{1}{n!~\I^n}H_n(\I x)=\frac 1{\I^n}\widetilde H_n(\I x)~~~\mbox{for $n\geq 0$} 
\end{aligned} 
\end{aligned}
\label{herm1}\ee
with $H_0=\widetilde H_0=P_0=1$ and
\be
H_n(x)=\tilde H_n(x)=P_n(x)=0
~~~~\mbox{for $n< 0$}.
\label{herm2}\ee
 The integral $\Phi_n$, given in (\ref{Phi}), has the following well-known form\footnote{See e.g. \cite{AvM}, section 4.1. }:
\be
\begin{aligned}
 \Phi_n (\eta) &:=\frac{1}{2\pi \I} \int_L \frac{e^{v^2+2\eta v}}{v^{n+1}} dv  
 =\frac{e^{-\eta^2}}{\sqrt{\pi}} 
 2^n\times \left\{\begin{aligned}
  &   
  \int^\infty_0 \frac {\xi^n} {n!} e^{- \xi^2+2\xi \eta} d\xi  \quad , n\quad \geq 0
\\  \\&
    H_{-n-1}(-\eta)  \quad ,\quad n \leq -1
 \end{aligned}\right. \\ 
  \end{aligned}
 \label{Phi-Psi}\ee


 \begin{lemma}\label{lemmaVol}
 The following holds for $0\leq \tau_1<\tau_2$:
 \be \bl\label{volformula}
\mbox{Vol}&~(\CR(\tau_1,x;\tau_2,y)) 
\\=&
\det\left(
\BH^{\tau_2-\tau_1}(y_i\!-\!x_1),\dots,
\BH^{\tau_2-\tau_1}(y_i\!-\!x_{n_1}),
\bar H_{n_2-n_1-1}(y_i),\dots,
\bar H_{0}(y_i)
\right)_{1\leq i\leq n_2}
%
\el \ee
Notice that for for $0\leq \tau_1<\tau_2\leq \rho$, the Hermite part is totally absent, since then $n_1=n_2=r$.
 \end{lemma}
  \begin{corollary}
For $\rho<\tau_1<\tau_2$, $\mbox{Vol}~(\CR(\tau_1,x;\tau_2,y))$ can also be written as:
 \be \bl\label{volformula'}
\mbox{Vol}&~(\CR(\tau_1,x;\tau_2,y)) 
\\=& (-1)^{\frac 12 (n_2-n_1)(n_2+n_1-1)}
\\&
~~ \det\left(\bar H_{0}(y_i),\dots, \bar H_{n_2-n_1-1}(y_i), \BH^{\tau_2-\tau_1}(y_i\!-\!x_1),\dots,
\BH^{\tau_2-\tau_1}(y_i\!-\!x_{n_1})\right)_{1\leq i\leq n_2}
  \el\ee
\end{corollary}
 \proof
 The proof proceeds by induction: at first, for $n_2=n_1+1$ and $\tau_2=\tau_1+1$, the right hand side of (\ref{volformula}) reads, using $\bar H_{0}(y_i)=1$ and setting $x_{n_1+1}=-\infty$,
$$
 \bl \det&\left(
\BH^{ 1}(y_i\!-\!x_1),\dots,
\BH^{ 1}(y_i\!-\!x_{n_1}),
1
\right)_{1\leq i\leq n_1+1}
\\&=
 \det\left(
\Id_{x_1\leq y_i } ,\dots,\Id_{x_{n_1}\leq y_i },
1
\right)_{1\leq i\leq n_1+1}
\\&=
 \det\left(
\Id_{x_1\leq y_i } ,\dots,\Id_{x_{n_1}\leq y_i },
\Id_{x_{n_1+1}\leq y_i }
\right)_{1\leq i\leq n_1+1}
\\& =\Id_{x\preceq y}= {\mbox{Vol} }(\CR(\tau_1,x;\tau_1+1,y)) .
\el$$
 Next, 
  given the formula (\ref{volformula}), we show its validity for $\tau_2\mapsto \tau_2+1$ and $n_2\mapsto n_2+1$. Indeed, setting $u=(u_{n_2+1},u_{n_2},\dots, u_1)\succ y$, the volume $\mbox{Vol}(\CR(n_1,x;n_2+1,u)) $ can be computed in terms of $\mbox{Vol}(\CR(n_1,x;n_2,y))$, which by the inductive step equals formula (\ref{volformula}):
 \be\bl\label{proofvol}
& \mbox{Vol} (\CR(\tau_1,x;\tau_2+1,u)) 
\\& =\int_{u_{n_2+1}}^{u_{n_2 }} dy_{n_2}\dots
\int_{u_{k+1 }}^{u_{k }} dy_{k} \dots
\int_{u_{2 }}^{u_{1 }} dy_{1} 
\mbox{Vol}(\CR(\tau_1,x;\tau_2,y)) 
\\&=\int_{u_{n_2+1}}^{u_{n_2 }} dy_{n_2}\dots
\int_{u_{k+1 }}^{u_{k }} dy_{k} \dots
\int_{u_{2 }}^{u_{1 }} dy_{1} 
\\&
~~~\det\left(
\BH^{\tau_2-\tau_1}(y_i\!-\!x_1),\dots,
\BH^{\tau_2-\tau_1}(y_i\!-\!x_{n_1}),
\bar H_{n_2-n_1-1}(y_i),\dots,
\bar H_{0}(y_i)
\right)_{1\leq i\leq n_2}
\\&
=\det\left(
\Bigl(\int_{u_{i+1 }}^{u_{i }} dy_{i}\BH^{\tau_2-\tau_1}(y_i\!-\!x_j)\Bigr)_{{1\leq i\leq n_2}\atop{1\leq j\leq n_1}} 
\Bigl(\int_{u_{i+1 }}^{u_{i }} dy_{i}\bar H_{n_2-j}(y_i)\Bigr)_{{1\leq i\leq n_2}\atop{n_1+1\leq j \leq n_2 }}  \right)
\\&
\stackrel{*}{=}\det\left(\Bigl(a_{ij}-a_{i+1,j}\Bigr)_{{1\leq i,j\leq n_2} } 
\right)=\det\left(\Bigl(a_{ij}
 \Bigr)_{{1\leq i\leq n_2+1}\atop{1\leq j\leq n_2}}\begin{array}{c}\bar H_0 \\ \vdots\\ \bar H_0\end{array}\right), \mbox{  since  }\bar H_0=1,
 \el\ee
 where in $\stackrel{*}{=}$ we use the identities below; noticing that in ({\ref{volformula}), we may replace momentarily $\bar H_n(x) \mapsto \bar{\bar{ H_n}}(x)=\frac{1}{n!}H_n(x/2)=\frac{x^n}{n!}+\dots$ without changing the volume. It has the advantage that $\bar{\bar{ H_n}}'(x)=\bar{\bar{ H}}_{n-1} (x)$; this will be used in the second formula below:
 $$ \bl
 \int_{u_{i+1}}^{u_i}dy_i\BH^{(\tau_2-\tau_1)}(y_i-x_j) &=\BH^{(\tau_2-\tau_1+1)} (u_i-x_j)\!\!-\!\!
\BH^{(\tau_2-\tau_1+1)} (u_{i+1}-x_j))
\\&=:a_{i,j}-a_{i+1,j}, \mbox{  for  }1\leq j\leq n_1,
 \\ \int_{u_{i+1 }}^{u_{i }} dy_{i}\bar{\bar{ H }}_{n_2- j}(y_i\!-\!x_{n_1})&=
\bar{\bar{ H }}_{n_2- j+1}(u_i)-
\bar{\bar{ H }}_{n_2- j+1}(u_{i+1}))
\\&=:a_{i,j}-
a_{i+1,j}, \mbox{  for  }n_1+1\leq j\leq n_2,
 \el$$
establishing the first formula of Lemma  \ref{lemmaVol} for $\rho\leq \tau_1\leq \tau_2$. The case $0\leq \tau_1\leq \tau_2\leq \rho$ proceeds along similar lines. As to the corollary and formula (\ref{volformula'}) : the $\bar H_i$ have to be permuted and passed though the $\BH^{(\tau_2-\tau_1+1)} (y_i-x_j)$'s; this produces $\tfrac 12 (n_2-n_1)(n_2-n_1-1)+n_1(n_2-n_1)=\frac 12 (n_2-n_1)(n_2+n_1-1)$ sign changes.\qed

%

 \newpage

 \section{Expressing the joint density as the product of two determinants}

 Instead of the kernel $ {\mathbb L}^{\mbox{\tiny dTac}}(\tau_1,\theta_1;\tau_2,\theta_2)$, it will be more convenient to consider the kernel $\widetilde {\mathbb L}^{\mbox{\tiny dTac}} $ in the  new variables $x $ and $y $, as mentioned in (\ref{theta-xy}):
\be\label{Ltilde}\bl
\widetilde {\mathbb L}^{\mbox{\tiny dTac}} 
 (\tau_1, x  ;\tau_2, y ) 
=    e^{\frac{y ^2}2}  {\mathbb L}^{\mbox{\tiny dTac}}
 (\tau_1, -2x  ;\tau_2, -2y )  e^{-\frac{x ^2}2}
\el\ee
 Given $n_i$ points at level $\tau_i$, the following probability can be expressed in terms of the discrete tacnode kernel $\widetilde {\mathbb L}^{\mbox{\tiny dTac}}$, given in (\ref{Ltilde}), as

\be \label{Prob}\bl\BP&\left(\begin{array}{l} \theta_{1,i} \in d\theta_{1,i}\mbox{, belonging to level $\tau_1$,  for } 1\leq i\leq n_1,\\ \theta_{2,i} \in d\theta_{2,i}\mbox{, belonging to level $\tau_2$, for } 1\leq j\leq n_2\end{array}\right)
\\&=
\det \left(
\begin{array}{cccccc}
\left(
  {\mathbb L}^{\mbox{\tiny dTac}}
 (\tau_1 , \theta_{1,i} ;\tau_1 , \theta_{1,j})\right)_{1\leq i,j\leq n_1}
&
\left(   {\mathbb L}^{\mbox{\tiny dTac}}
 (\tau_1 , \theta_{1,i} ;\tau_2 , \theta_{2,j})\right)_{{1\leq i \leq n_1}\atop{1\leq  j\leq n_2}}
 \\
 \left(  {\mathbb L}^{\mbox{\tiny dTac}}
 (\tau_2 , \theta_{2,i} ;\tau_1 , \theta_{1,j})\right)_{{1\leq i \leq n_2}\atop{1\leq j\leq n_1}}
&
\left(   {\mathbb L}^{\mbox{\tiny dTac}}
 (\tau_2 , \theta_{2,i} ;\tau_2,\theta_{2,j}  )\right)_{{1\leq i,j\leq n_2} }
\end{array}\right)
\\&~~\hspace{7cm}~~~\prod_{i=1}^{n_1} d \theta_{1,i}  \prod_{j=1}^{n_2} d \theta_{2,j}
\\&=\det \left(
\begin{array}{cccccc}
\left(
\widetilde {\mathbb L}^{\mbox{\tiny dTac}}
 (\tau_1 , x_i ;\tau_1 , x_j)\right)_{1\leq i,j\leq n_1}
&
\left( \widetilde {\mathbb L}^{\mbox{\tiny dTac}}
 (\tau_1 , x_i ;\tau_2 , y_j)\right)_{{1\leq i \leq n_1}\atop{1\leq  j\leq n_2}}
 \\
 \left(\widetilde {\mathbb L}^{\mbox{\tiny dTac}}
 (\tau_2 , y_i ;\tau_1 , x_j)\right)_{{1\leq i \leq n_2}\atop{1\leq j\leq n_1}}
&
\left( \widetilde {\mathbb L}^{\mbox{\tiny dTac}}
 (\tau_2 , y_i ;\tau_2 , y_j)\right)_{{1\leq i,j\leq n_2} }
\end{array}\right)
\\&~~\hspace{7cm}~~\Bigr|(-2)^{n_1+n_2}\Bigr|~\prod_{i=1}^{n_1}  dx_i 
\prod_{j=1}^{n_2}  dy_j 
\\& =: p(\tau_1,{\bf x};\tau_2,{\bf y})\prod_{i=1}^{n_1}  dx_i 
\prod_{j=1}^{n_2}  dy_j   
\el\ee
with density $p(\tau_1,{\bf x};\tau_2,{\bf y})$.

 \begin{proposition}\label{Prop3.1}
For $\tau_1,\tau_2\geq \rho$, we have
  \be\bl 
 \widetilde {\mathbb L}^{\mbox{\tiny dTac}}&
 (\tau_1, x  ;\tau_2, y ) 
 \\
= &  e^{\frac{y ^2}2}  {\mathbb L}^{\mbox{\tiny dTac}}
 (\tau_1, -2x  ;\tau_2, -2y )  e^{-\frac{x ^2}2}
\\
=& \frac {1}{\sqrt{\pi}}\sum_{\al=\tau_1-\rho}^{\tau_1-\rho+r-1}
e^{-\frac{x^2}2}G^{(\tau_1)}_{\al}(x)
 e^{-\frac{y^2}2}\widehat H_{\al -\tau_1+\tau_2 }(y)
 \\&
 +\frac{1}{\sqrt{\pi}}\sum_{\al=\max(0,\tau_1-\tau_2)}^{\tau_1-\rho-1}
 e^{-\frac{x^2}2}\widetilde H_{\al }(x )
 e^{-\frac{y^2}2}\widehat H_{\al -\tau_1+\tau_2 } (y)
\\&+\Id_{\tau_1>\tau_2}e^{-\frac{x^2}2}\left(-{\mathbb H}^{\tau_1-\tau_2}( 2(x-y) 
 ) +
 \sum_{\al=0}^{\tau_1-\tau_2 -1}
 \widetilde H_{\al}(x )
 \Phi_{\tau_1-\tau_2 -\al-1}(-y)
\right)e^{ \frac{y^2}2}
\el\label{Ltacnode}\ee
where
\be\bl
\Ga_{\ell,k}^{(r)}(\beta):=\frac{(-1)^{\ell+1}}{\Theta_r( 0,0)}
\left[     
\prod_{\al=1}^r\oint_{\uparrow L_{0+}}\frac{e^{    2w_\al^2+\beta   w_\al}}{    w_\al  ^{\rho+1 }}
 \frac{dw_\al}{2\pi \I}\right]
  { \Dt_r^2(w)}{}
 \sg_{r-\ell-1}^{(r)}(w) h_k(w ^{-1}) 
 \el\label{Gamma}\ee
 $$\bl
\widetilde\Ga_{\ell,k}^{(r-1)}(\beta):=\frac{ (-1)^{\ell+k}r}{\Theta_r( 0,0)}
\left[     
\prod_{\al=1}^{r-1}\oint_{\uparrow L_{0+}}\frac{e^{    2w_\al^2+\beta   w_\al}}{    w_\al  ^{\rho  }}
 \frac{dw_\al}{2\pi \I}\right]
    \Dt_{r-1}^2(w)
 \sg_{r-\ell-1}^{(r-1)}(w) 
 \sg_{r-k-1}^{(r-1)}(w) ,
 \el$$
 and
 $$
 G^{(\tau_1)}_{\tau_1-\rho+k}(x):= \sum_{i=0}^{\tau_1-\rho  -1}C_{k,i}
\widetilde H_{ i}(x)+ \sum_{i=0}^{   r-1} \widetilde \Ga_{k,i }^{(r-1)}(\beta)
 \Phi_{\tau_1-i-1}(x+\tfrac{\beta }{2}) 
 $$
 with
 \be \label{C}C_{k,\al}:=\sum_{\ell'=0 }^{\min(\tau_1-\rho-\al-1,r-k-1)}  \Ga^{(r)}_{\ell'+k,\tau_1-\rho- \ell'-\al-1}(\beta).
 \ee
 
 \end{proposition}

\proof For $\tau_1,\tau_2\geq \rho$, the only surviving terms in (\ref{Final0}) are ${\mathbb K}^{\mbox{\tiny GUE}}$ and $ {\mathbb L}^{\mbox{\tiny dTac}}_i  (\tau_1, \theta_1 ;\tau_2, \theta_2)  $ for $i=1,2$.

\medbreak

\noindent {\bf (i) Expression for $ {\mathbb K}^{\mbox{\tiny GUE}}   ( \tau_1\!-\!\rho, -\theta_1 ;\tau_2\!-\!\rho, -\theta_2)$.} 
Using the formulas (\ref{herm1}) and (\ref{Phi-Psi}), we have for the GUE-kernel ${\mathbb K}^{\mbox{\tiny GUE}}$:\footnote{
 In $\stackrel{*}{ =}$, when $\tau_1\leq \tau_2$, then $\Phi$'s subscript $\rho-\tau_2+j\leq -1$ for all $0\leq j\leq \tau_1-\rho-1$.}
\be
\bl
{\mathbb K}^{\mbox{\tiny GUE}}&  ( \tau_1\!-\!\rho, -\theta_1 ;\tau_2\!-\!\rho, -\theta_2)  +  
{\mathbb H}^{\tau_1-\tau_2}(  \theta_2-  \theta_1)
\\&=
\sum_{j=0}^{\tau_1-\rho-1}\oint_{\Ga_0}\frac{du}{2\pi \I}\frac{e^{-u^2-\theta_1 u}}{u^{\tau_1-\rho-j }}
\oint_{ L_{0+}} \frac{dv}{2\pi \I} \frac{e^{ v^2+\theta_2 v}}{v^{\rho-\tau_2+j+1}}
\\&\stackrel{*}{ =}
 (\Id_{\tau_1\leq \tau_2}+\Id_{\tau_1> \tau_2})
\sum_{j=0}^{\tau_1-\rho-1}\widetilde 
H_{\tau_1-\rho-j-1}(-\tfrac{\theta_1}{2})
\Phi_{\rho-\tau_2+j}(\tfrac{\theta_2}{2})
\\&=
 \frac{e^{-\frac {\theta_2^2}{4}}}{\sqrt{\pi}}
\sum_{j=\rho+1}^{\min(\tau_1,\tau_2)}\widetilde 
H_{\tau_1 -j }(-\tfrac{\theta_1}{2})
 \widehat H_{\tau_2 -j } (-\tfrac{\theta_2}{2})
\\&\qquad \qquad \qquad \qquad \quad  +\Id_{\tau_1>\tau_2}\sum_{j=\tau_2}^{\tau_1 -1}\widetilde 
H_{\tau_1 -j-1}(-\tfrac{\theta_1}{2})
\Phi_{j-\tau_2 }( \tfrac{\theta_2}{2}).
%
\el
\label{KGUE}\ee
\medbreak
\noindent {\bf (ii) Expressions for $ {\mathbb L}^{\mbox{\tiny dTac}}_i  (\tau_1, \theta_1 ;\tau_2, \theta_2)  $ for $i=1,2$.} At first we express the polynomial  $P_r^{(w)}(z)$ of degree $r$ (defined below) and its inverse in terms of symmetric functions,
  $$\bl P_r^{(w)}(z)&:=\prod_1^r(z-w_\al)=\sum_{\ell=0}^r
(-1)^{r-\ell}  \sg_{r-\ell}^{(r)}(w)z^\ell
\\
P_r^{(w)}(z)^{-1}&=\frac{1}{\prod_1^r (-w_\al)}\prod_{\al=1}^r\frac{1}{1-\frac z{w_\al}}=
\frac{(-1)^r}{\prod_1^r  w_\al }\sum_{k=0}^\infty h_k(w ^{-1})z^k,
\el $$
and thus
$$\bl
\frac{P^{(w)}_r(v)-P^{(w)}_r(u)}{(v-u)P^{(w)}_r(u)} = \frac{1}{\prod_1^r w_\al}\sum_{\ell=0}^{r-1} \sum_{k=0}^\infty(-1)^{ \ell+1}\sg_{r-\ell-1}^{(r)}(w) h_k(w ^{-1}) 
\sum_{i+j=\ell}v^iu^{j+k}.
\el$$
This is used in the formula for ${\mathbb L}^{\mbox{\tiny dTac}}_i  $ for $i=1,2$; indeed:
$$
\bl
\frac{\Theta_r( u, v)-\Theta_r( 0,0)}{(v-u)\Theta_r( 0,0)}&:=\left[     
\prod_{\al=1}^r\oint_{\uparrow L_{0+}}\frac{e^{    2w_\al^2+\beta   w_\al}}{    w_\al  ^{\rho }}
 \frac{dw_\al}{2\pi \I}\right]
 \left(\frac{P^{(w)}_r(v)-P^{(w)}_r(u)}{(v-u)P^{(w)}_r(u)} \right) 
\frac{ \Dt_r^2(w)}{\Theta_r( 0,0)}
\\
&=\sum_{\ell=0}^{r-1}  \sum_{i+j=\ell}\sum_{k=0}^\infty v^iu^{j+k}\Ga^{(r)}_{\ell,k}(\beta),
\el
$$
$$\bl
r\frac{ \Theta^+_{r-1}(  u, v )} {  \Theta_r(0,0)} 
  &:=\frac1{  \Theta_r(0,0)} \left[     
\prod_1^{r-1}\oint_{\uparrow L_{0+}}\frac{e^{    2w_\al^2+\beta  w_\al}}{    w_\al  ^{\rho }}
 ~\left( ({v\!-\!w_\al} )\ ({u\!-\!w_\al})\right)\frac{dw_\al}{2\pi \I}\right]\Dt_{r-1}(w)
\\&=\sum_{0\leq k,\ell\leq r-1}
v^k u^\ell\widetilde \Ga^{(r-1)}_{k,\ell}(\beta)
\el $$
where $\Ga_{\ell,k}^{(r)}(\beta)$ and $\widetilde\Ga_{\ell,k}^{(r-1)}(\beta)$ are defined in
(\ref{Gamma}).
 
Using these expressions and referring to $C_{k,\al}$ as in (\ref{C}), we have
\be\bl
{\mathbb L}^{\mbox{\tiny dTac}}_1& (\tau_1, \theta_1 ;\tau_2, \theta_2) 
\\&= \oint_{\Ga_0}\frac{du} {(2\pi\I)^2}\oint_{\uparrow L_{0+}}    dv \frac{v^{\tau_2-\rho }}{u^{\tau_1-\rho }}
\frac{e^{-u^2  -  \theta_1    u  }}
{e^{-v^2 - \theta_2   v  }}
      \frac{ \Theta_r( u, v )-\Theta_r(0,0)} {(v-u) \Theta_r(0,0)}  
\\&=
  \sum_{\ell=0}^{r-1}  \sum_{i+j=\ell}\sum_{k=0}^\infty
  \oint_{\Ga_0}\frac{du} {(2\pi\I)^2}\oint_{\uparrow L_{0+}}  dv\frac{v^{\tau_2-\rho+i }}{u^{\tau_1-\rho-j-k }}
\frac{e^{-u^2  -  \theta_1    u  }}
{e^{-v^2 - \theta_2   v  }}
   \Ga^{(r)}_{\ell,k}(\beta)
\\ &= \frac {e^{-\frac{\theta_2^2}4}}{\sqrt{\pi}}\sum_{\ell=0}^{r-1}\sum_{i+j=\ell} \sum_{k=0}^{\tau_1-\rho-j-1}
  \widetilde H_{ \tau_1-\rho-j-k-1}(-\tfrac{\theta_1}2)
\widehat H_{\tau_2-\rho+i}(-\tfrac{\theta_2}2)
 \Ga^{(r)}_{\ell,k}(\beta)
\\ &= \frac {e^{-\frac{\theta_2^2}4}}{\sqrt{\pi}}\sum_{k=0}^{ r-1} 
\widehat H_{\tau_2-\rho+k}(-\tfrac{\theta_2}2)
 \sum_{\ell=k}^{r-1}~~\sum_{\al=0}^{\tau_1-\rho-(\ell-k)-1}
  \widetilde H_{ \al}(-\tfrac{\theta_1}2)
 \Ga^{(r)}_{\ell,\tau_1-\rho-(\ell-k)-\al-1}(\beta)
 \\ &= \frac {e^{-\frac{\theta_2^2}4}}{\sqrt{\pi}}\sum_{k=0}^{ r-1} 
\widehat H_{\tau_2-\rho+k}(-\tfrac{\theta_2}2)
\\&~~~~~~\times~~ \sum_{\ell=k}^{\min(r-1, \tau_1-\rho+k-1)}~~\sum_{\al=0}^{\tau_1-\rho-(\ell-k)-1}
  \widetilde H_{ \al}(-\tfrac{\theta_1}2)
 \Ga^{(r)}_{\ell,\tau_1-\rho-(\ell-k)-\al-1}(\beta)
 \\ &=  \frac {e^{-\frac{\theta_2^2}4}}{\sqrt{\pi}}\sum_{k=0}^{ r-1} 
\widehat H_{\tau_2-\rho+k}(-\tfrac{\theta_2}2)
\\&~~~~~~\times~~\sum_{\al=0}^{\tau_1-\rho  -1}
\widetilde H_{ \al}(-\tfrac{\theta_1}2)
\sum_{\ell'=0 }^{\min(\tau_1-\rho-\al-1,r-k-1)}  \Ga^{(r)}_{\ell'+k,\tau_1-\rho- \ell'-\al-1}(\beta)
 \\ &=: \frac {e^{-\frac{\theta_2^2}4}}{\sqrt{\pi}}\sum_{k=0}^{ r-1} 
\widehat H_{\tau_2-\rho+k}(-\tfrac{\theta_2}2)
\sum_{\al=0}^{\tau_1-\rho  -1}C_{k,\al}
\widetilde H_{ \al}(-\tfrac{\theta_1}2),
 \el\label{L1}\ee
 %
%

\newpage

\vspace*{-2cm}
\noindent and
\be\bl
 {\mathbb L}^{\mbox{\tiny dTac}}_2 (&\tau_1, \theta_1 ;\tau_2, \theta_2) 
 \\ &=
  \oint_{\uparrow L_{0+}  }    \frac{du} {(2\pi\I)^2} \oint_{\uparrow L_{0+}}dv  \frac{v^{\tau_2-\rho}}{u^{\tau_1}}
\frac{e^{ u^2 -    (\theta_1- \beta     )u  }}
{e^{-v^2 - \theta_2    v  }}
   ~ \frac{ r\Theta^+_{r-1}(  u, v )} {  \Theta_r(0,0)} 
    \\&=
\sum_{0\leq k,\ell \leq r-1}\widetilde \Ga_{k,\ell}^{(r-1)}(\beta)\oint_{  L_{0+}  }    \frac{du} {(2\pi\I)^2} \oint_{  L_{0+}}dv  \frac{v^{\tau_2-\rho+k}}{u^{\tau_1-\ell}}
\frac{e^{ u^2 -    (\theta_1- \beta     )u  }}
{e^{-v^2 - \theta_2    v  }}
 \\
 &=\sum_{0\leq k,\ell \leq r-1}
 \widetilde \Ga_{k,\ell }^{(r-1)}(\beta)
 \Phi_{\tau_1-\ell-1}(\tfrac{\beta-\theta_1}{2})
  \Phi_{\rho-\tau_2-k-1}(\tfrac{ \theta_2}{2})
  \\
 &\stackrel{(\tau_2\geq \rho)}{=} \frac {e^{-\frac{\theta_2^2}4}}{\sqrt{\pi}}\sum_{k=0}^{r-1}\widehat H_{\tau_2-\rho+k}(-\tfrac{ \theta_2}{2})
 \sum_{ 0\leq  \ell \leq r-1} \widetilde \Ga_{k,\ell }^{(r-1)}(\beta)
 \Phi_{\tau_1-\ell-1}(\tfrac{\beta-\theta_1}{2}).
\label{L2} \el \ee
Adding the three formulas (\ref{KGUE}), (\ref{L1}) and (\ref{L2}), setting $\theta_1=-2x$ and $\theta_2=-2y$, changing the summing index to $\al=\tau_1-j-1$ in (\ref{KGUE}) and $\al=\tau_1-\rho+k $ in both (\ref{L1}) and (\ref{L2}) and conjugating by $e^{-\frac{ x^2}2}$ and $e^{\frac{y^2}2}$, we find that formula (\ref{Final0}) becomes (\ref{Ltacnode}) for $\tau_1,\tau_2\geq \rho$, ending the proof of proposition \ref{Prop3.1}.\qed

   
 \begin{proposition}\label{Prop3.2} For $0\leq\tau_1<\tau_2$, the two-level density is given by 
\be \label{2-kernel'}\bl
 \frac{1}{2^{n_1+n_2}}&p(\tau_1,{\bf x};\tau_2,{\bf y}) 
%
 \\&=\det\left(
   \begin{array}{cccc}( A_1^\top(x_i) B_1(x_j))_{1\leq i,j\leq n_1} &
   (A_1^\top(x_i) B_2(y_j))_{{1\leq i \leq n_1}\atop{1\leq  j\leq n_2}}
 \\
 (A_2^\top(y_i) B_1(x_j))_{{1\leq i \leq n_2}\atop{1\leq j\leq n_1}} & ( A_2^\top(y_i) B_2(y_j))_{1\leq i,j\leq n_2}
 \end{array} \right)
 \\
 &=\det \left(\left\la\AR_\al,  {\mathcal B}_\beta\right\ra\right)_{1\leq \al,\beta\leq n_1+n_2}
 =\det\left( \AR \right)
 \det\left( {\mathcal B} \right)
 \el
 \ee
where  $A_i$ and $B_i\in \BC^{n_1+n_2}$ are column-vectors  and where $\AR:=(\AR_\al)_{1\leq \al\leq n_1+n_2}$ is a column of row-vectors and 
${\mathcal B}=({\mathcal B}_\beta)_{1\leq \beta\leq n_1+n_2}$ a row of column-vectors (both ${\mathcal A}$ and ${\mathcal B}$ can be interpreted as square matrices of size $n_1+n_2$):
\be\label{ABscrip}\AR 
 :=\left(\begin{array}{ccc}
 A_1^\top (x_1) \\ \vdots \\ A_{1}^\top (x_{n_1})\\ \\ A_2^\top (y_{1})\\ \vdots\\ A_2^\top (y_{ n_2})
 \end{array}\right)
~\mbox{and}~ {\mathcal B}  
 =\left(B_1(x_1),\dots,B_1(x_{n_1}) ,
 B_2(y_1),\dots,B_2(y_{n_2}) \right).
 \ee
  \noindent{\bf case 1: $\rho<\tau_1<\tau_2$}: here the $A_i$ and $B_i$ are as follows:
\end{proposition}
\newpage
 
 
\newpage

  \setlength{\unitlength}{0.017in}\begin{picture}(0,60)
\put(115,-70){\makebox(0,0) {\rotatebox{0}{\includegraphics[width=185mm,height=216mm]
 {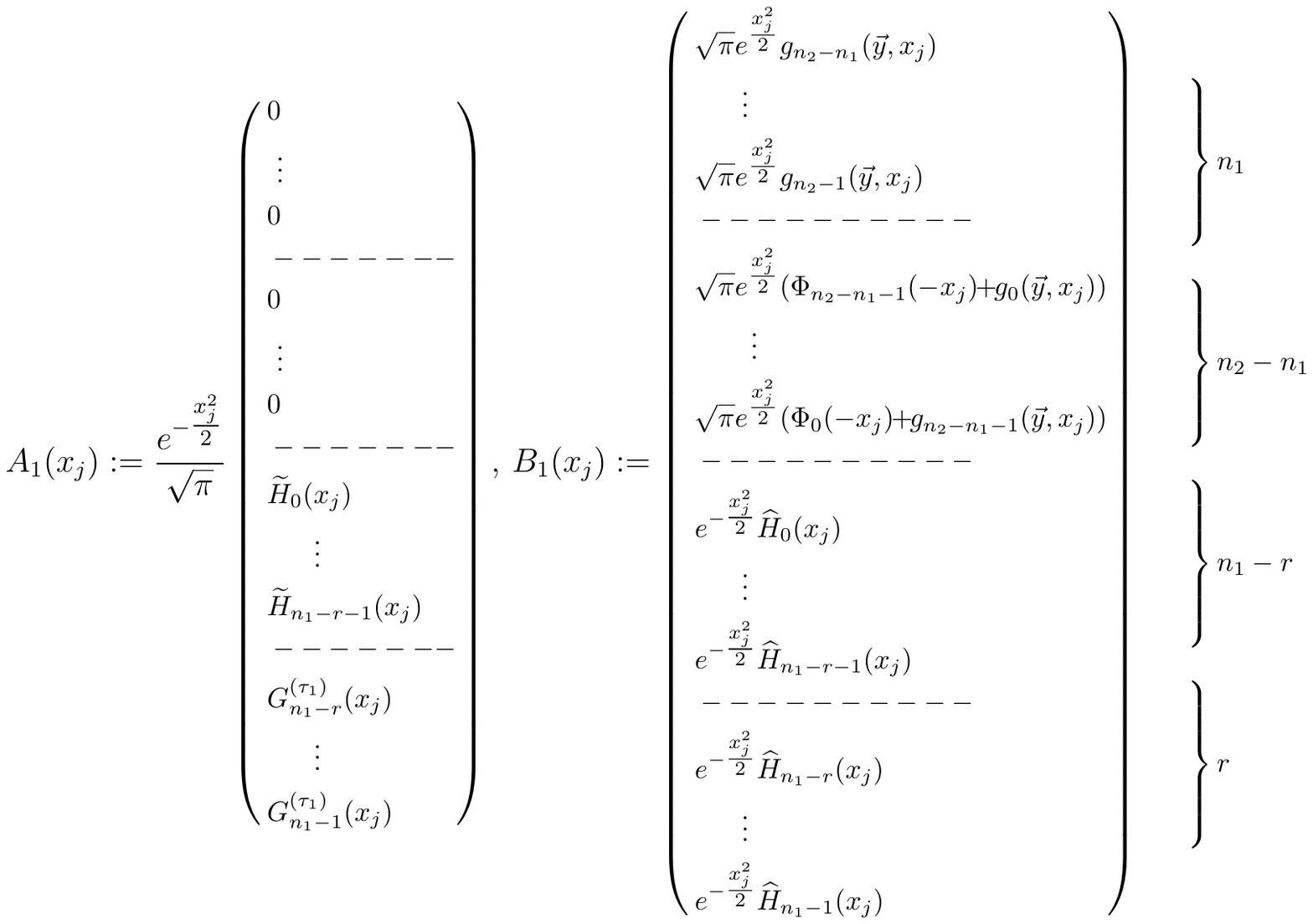}}}}

\put(135,-300){\makebox(0,0) {\rotatebox{0}{\includegraphics[width=220mm,height=216mm]
 {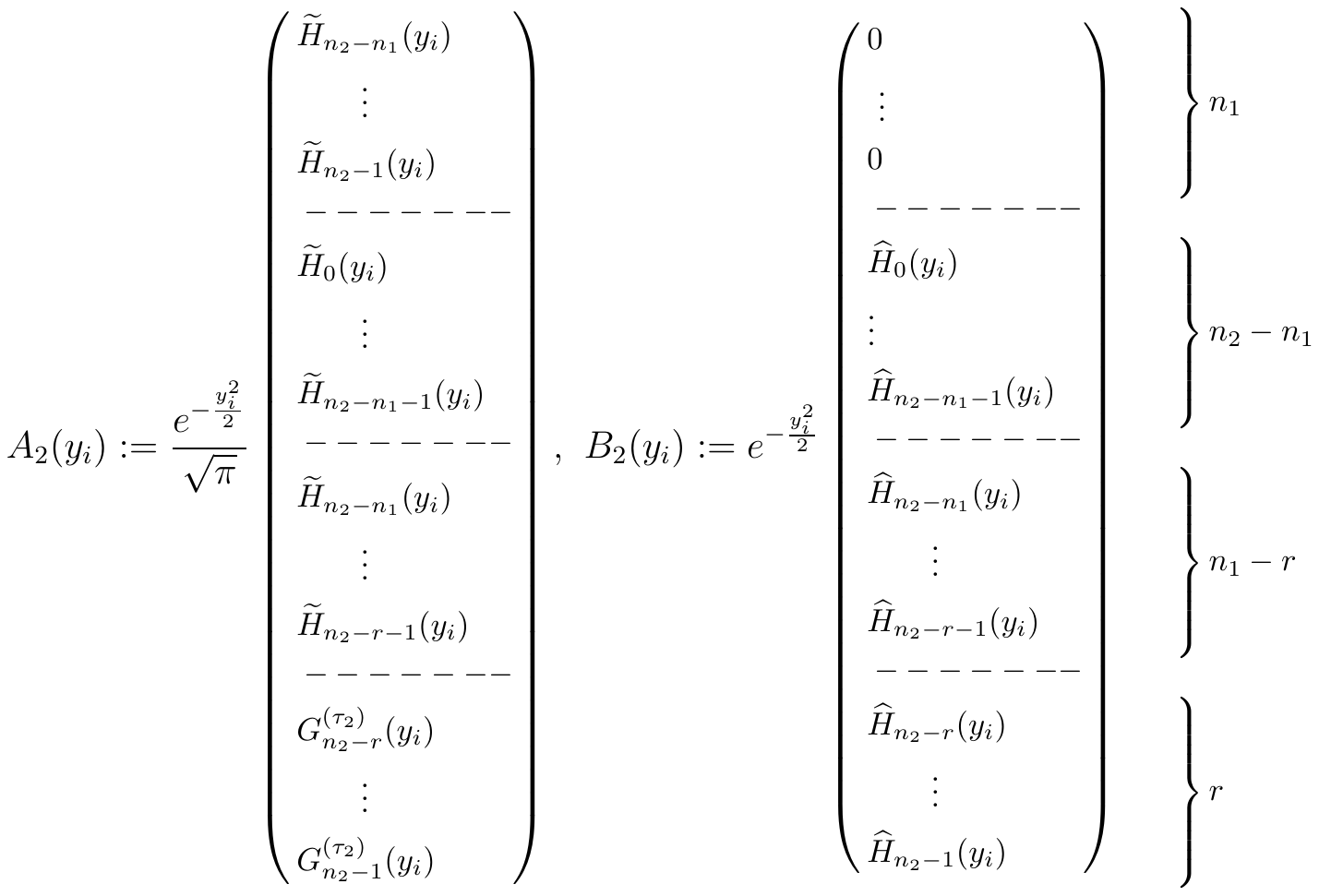}}}}
 
   \end{picture}
   
   \vspace*{15cm}
   
   \be\label{AB}\ee

\newpage

 \noindent{\bf case 2: $0\leq\tau_1<\tau_2\leq \rho$}: here the $A_i$ and $B_i$ are as follows:

$$\footnotesize
A_1(x_i)\!=\!
e^{-\frac{x_i^2}2}\left(
\begin{array}{ccc}
0
\\ \vdots
\\ 0
\\ \Phi_{\tau_1-1}(x_i\!+\!\tfrac\beta 2)
\\ \vdots
\\ \Phi_{\tau_1-r}(x_i\!+\!\tfrac\beta 2)
\end{array}
\right)
~,~~
B_1(x_j)\!=\!e^{ \frac{x_j^2}2}
\left(
\begin{array}{ccc}
g_0(\vec y,x_j)
\\ \vdots
\\ g_{r-1} (\vec y,x_j)
\\ {\dis\sum_{k=0}^{r-1}}\widetilde \Ga_{k,0}\Phi_{\rho-\tau_1-k-1}(-x_j )
\\ \vdots
\\ {\dis\sum_{k=0}^{r-1}}\widetilde \Ga_{k,r-1}\Phi_{\rho-\tau_1-k-1}(-x_j )
\end{array}
\right)
\hspace*{-.9cm}
\bl &\left.\begin{array}{ll}&\\&\\&\\ \end{array}\right\}r\\  \\
&\left.\begin{array}{ll}&\\&\\&\\&\\&\\ \end{array}\right\}r
\el
$$
\be\footnotesize
A_2(y_i)\!=\!
e^{-\frac{y_i^2}2}\left(
\begin{array}{ccc}
\widetilde H_0(y_i)
\\ \vdots
\\ \widetilde H_{r-1}(y_i)
\\   \Phi_{\tau_2-1}(y_i\!+\!\tfrac\beta 2)
\\ \vdots
\\ \Phi_{\tau_2-r}(y_i\!+\!\tfrac\beta 2)
\end{array}
\right)
~,~~
B_2(y_j)\!=\!e^{ \frac{y_j^2}2}
\left(
\begin{array}{ccc}
0
\\ \vdots
\\ 0
\\ {\dis\sum_{k=0}^{r-1}}\widetilde \Ga_{k,0}\Phi_{\rho-\tau_2-k-1}(-y_j )
\\ \vdots
\\ {\dis\sum_{k=0}^{r-1}}\widetilde \Ga_{k,r-1}\Phi_{\rho-\tau_2-k-1}(-y_j )
\end{array}
\right)
\hspace*{-.9cm}
\bl &\left.\begin{array}{ll}&\\&\\&\\ \end{array}\right\}r\\  \\
&\left.\begin{array}{ll}&\\&\\&\\&\\&\\ \end{array}\right\}r
.\el
\label{AB'}\ee

 \proof The different kernels in (\ref{2-kernel}) will be expressed in terms of inner-products involving $A_i$ and $B_i$, as in (\ref{AB}):
 
 \noindent {\bf Case 1: $\rho<\tau_1<\tau_2$}
 
 \noindent{\bf (i)  Expression of $\widetilde {\mathbb L}^{\mbox{\tiny dTac}}
 (\tau_1, x_i ;\tau_2, y_j)$ for $0\leq \rho<\tau_1\leq \tau_2$}. From (\ref{Ltacnode}), after setting $n_i-r=\tau_i-\rho$, we have 
\be\bl 
 \widetilde {\mathbb L}^{\mbox{\tiny dTac}}
 &(\tau_1, x_i ;\tau_2, y_j) 
\\ &\stackrel{*}{=} \frac{1}{\sqrt{\pi}}
\left[\bl
&\sum_{\al=0}^{n_1-r-1} 
e^{- \frac {x_i^2}{2}}
\widetilde H_{ \al}(x_i) 
e^{- \frac {y_j^2}{2}}
\widehat H_{\al-n_1+n_2 } (y_j)
\\&+\sum_{\al=n_1-r}^{n_1 -1} 
 e^{- \frac {x_i^2}{2}}G^{(\tau_1)}_{ \al}(x_i)
 e^{- \frac {y_j^2}{2}}\widehat H_{\al-n_1+n_2 } (y_j)
\el\right] 
\\&=\left\{\bl
 &   A^\top_1(x_i) B_2(y_j)  \mbox{, after setting  }\rho<\tau_1<\tau_2 \mbox{ in }\stackrel{*}{=},
\\&   A_1^\top(x_i) B_1(x_j)  \mbox{, after setting}\left\{\!\begin{array}{l}\rho<\tau_1=\tau_2 \\
r<n_1=n_2\end{array}\!\right\}\mbox{   and   }y_j=x_j \mbox{ in }\stackrel{*}{=}.\el\right.
\el  \label{A1B1}\ee
 
\medbreak
\noindent{\bf (ii)  Expression of $\widetilde {\mathbb L}^{\mbox{\tiny dTac}}
 (\tau_2, y_i ;\tau_1, x_j)$ also for $0\leq \rho<\tau_1\leq \tau_2$}. 
  The Heaviside function, present in this case, needs to be expressed in terms of Hermite polynomials (\ref{herm1}); this is done by solving the following linear system of $n_2$ equations in $n_2$ unknowns $g_0(\vec y,x),\dots,g_{n_2-1}(\vec y,x) $; the sum can then be split into two parts:
\be\bl
-{\mathbb H}^{\tau_2-\tau_1}( &2(y_i-x))
\\&=\sum_{\al=0}^{n_2-1}
\widetilde H_{ \al}(y_i)g_\al( \vec y,x)
\mbox{   for  }
1\leq i\leq n_2,
\\
 \label{Heavilin}  
&= \sum_{\al=0}^{n_2-n_1-1} 
\widetilde H_{ \al}(y_i)g_\al( \vec y,x)
+\sum_{\al=0}^{ n_1-1} 
\widetilde H_{ n_2-n_1+\al}(y_i)g_{ n_2-n_1+\al}( \vec y,x).
\el\ee
So, for  $\rho< \tau_1\leq \tau_2$, we have, using throughout $\tau_i-\rho=n_i-r$,
$$\bl
&\widetilde {\mathbb L}^{\mbox{\tiny dTac}}
 (\tau_2, y_i ;\tau_1, x_j) 
\\& \stackrel{**}{=}   \frac{1}{\sqrt{\pi}}
\left[
\bl 
&+\Id_{n_2>n_1}\sum_{\al=0}^{n_1-1}
 e^{-\frac{y_i^2}2}\widetilde H_{n_2-n_1+\al}(y_i) \sqrt{\pi}{e^{ \tfrac{ x_j^2}2}} g_{n_2-n_1+\al}(\vec y, x_j)
 \\&+\Id_{n_2>n_1}\sum_{\al=0}^{n_2-n_1-1}
  e^{-\frac{y_i^2}2}\widetilde H_{\al}(y_i)\sqrt{\pi}e^{\tfrac{x_j^2}2}(\Phi_{n_2-n_1-\al-1}(-x_j)
  +g_\al(\vec y,x_j))
\\&+
 \sum_{\al=0}^{n_1-r-1}
 e^{-\frac{y_i^2}2}\widetilde H_{n_2-n_1+\al}(y_i) {e^{-\tfrac{ x_j^2}2}} \widehat H_{\al}(x_j)
 \\
& +\sum_{\al=n_1-r}^{n_1-1}
 e^{-\tfrac{  y_i^2}2}   G_{n_2-n_1+\al}^{(\tau_2 )}(y_i) {e^{-\tfrac{ x_j^2}2}} \widehat H_{\al}(x_j)
 \el
\right]
\\&=\left\{\bl
 &    A^\top_2(y_i) B_1(x_j) \mbox{, after setting   }r<n_1<n_2 \mbox{ in } \stackrel{**}{=} ,
\\&  A^\top_2(y_i) B_2(y_j) \mbox{, after setting   }\left\{\begin{array}{l}\rho<\tau_1=\tau_2 \\
r<n_1=n_2\end{array}\right\}\mbox{   and   }x_j=y_j, \mbox{ in } \stackrel{**}{=}.\el\right.
\el $$

\noindent{\bf Case 2: $0\leq \tau_1<\tau_2\leq \rho$}

For $0\leq \tau_1\leq \tau_2\leq \rho$, the only surviving terms in $\widetilde {\mathbb L}^{\mbox{\tiny dTac}}$, as in (\ref{Final0}),
 are the Heaviside part and the term ${\mathbb L}^{\mbox{\tiny dTac}}_2$; so
$$ \bl
{\mathbb L}^{\mbox{\tiny dTac}}   ( \tau_1, -2x   ;\tau_2, -2y)  
=- 
{\mathbb H}^{\tau_1-\tau_2}(   2(x-y)  ) 
+{\mathbb L}^{\mbox{\tiny dTac}}_2 (\tau_1, -2x ;\tau_2, -2y).
\el$$
The Heaviside-term vanishes for all entries in (\ref{Prob}), except for the $\widetilde {\mathbb L}^{\mbox{\tiny dTac}}
 (\tau_2 , y_i ;\tau_1 , x_j)$ -entry; there we use the same expansion of the Heaviside-term in Hermite polynomials, as in (\ref{Heavilin}),
\be\label{Heavilin'}\bl
-{\mathbb H}^{\tau_2-\tau_1}( 2(x_i-y))
&=\sum_{\al=0}^{r-1}
\widetilde H_{ \al}(x_i)g_\al( \vec x,y)
\mbox{   for  }
1\leq i\leq r,
\el \ee
From (\ref{L2}), we deduce
\be\bl
\widetilde{\mathbb L}^{\mbox{\tiny dTac}}   ( &\tau_1,  x_i   ;\tau_2,  y_j)
\\&=e^{-\frac{x_i^2}{2}}{\mathbb L}^{\mbox{\tiny dTac}}    ( \tau_1, -2x_i   ;\tau_2, -2y_j)
e^{ \frac{ y_j^2}{2}}  
\\
 &\stackrel{*}{=}\sum_{\al=0}^{r-1}e^{-\frac{x_i^2}{2}}\Phi_{\tau_1-\al-1}
 (x_i+\tfrac \beta  2)e^{ \frac{ y_j^2}{2}}  \sum_{k=0}^{r-1}\widetilde \Ga_{k,\al}\Phi_{\rho-\tau_2-k-1}(-y_j)
\\&=\left\{\bl
 &   A^\top_1(x_i) B_2(y_j)  \mbox{  setting   }0\leq\tau_1<\tau_2\leq \rho \mbox{  in }\stackrel{*}{=} \mbox{ above},
\\&   A_1^\top(x_i) B_1(x_j)  \mbox{ setting   }0\leq \tau_1=\tau_2 \leq \rho\mbox{   and   }y_j=x_j\mbox{  in }\stackrel{*}{=} \mbox{ above},\el\right.
\el \label{A1B1'}\ee
and
$$\bl
\widetilde{\mathbb L}^{\mbox{\tiny dTac}}   ( &\tau_2,  y_i   ;\tau_1,  x_j)
\\&=e^{-\frac{y_i^2}{2}}{\mathbb L}^{\mbox{\tiny dTac}}    ( \tau_2, -2y_i   ;\tau_1, -2x_j)  
e^{ \frac{x_j^2}{2}}
\\
 &\stackrel{**}{=}e^{-\frac{y_i^2}{2}}\left(\Id_{\tau_1<\tau_2} \sum_{\al=0}^{r-1}\widetilde H_\al(y_i)g_\al(\vec y,x_j)\right.\\
 &+\left.\sum_{\al=0}^{r-1}\Phi_{\tau_2-\al-1}
 (y_i+\tfrac \beta  2)\sum_{k=0}^{r-1}\widetilde \Ga_{k,\al}\Phi_{\rho-\tau_1-k-1}(-x_j)
 \right)e^{ \frac{x_j^2}{2}}
\\&=\left\{\bl
 &   A^\top_2(y_i) B_1(x_j)  \mbox{  setting   }0\leq\tau_1<\tau_2\leq \rho \mbox{  in }\stackrel{**}{=} \mbox{ above},
\\&   A_2^\top(y_i) B_2(y_j)  \mbox{   setting   }0\leq \tau_1=\tau_2 \leq \rho\mbox{   and   }x_j=y_j \mbox{  in }\stackrel{**}{=} \mbox{ above}.\el\right.
\el $$
ending the proof of Proposition \ref{Prop3.2}.\qed

\begin{proposition}\label{Prop4.3} For $0\leq \tau$, the one-level density is given by 
\be \label{2-kernel}\bl
 \frac{1}{2^{ n }} p(\tau ,{\bf x} ) 
%
  &=\det\left(
   ( \bar A_1^\top(x_i) \bar B_1(x_j))_{1\leq i,j\leq n } 
   \right)
 \\
 &=\det \left(\left\la\bar\AR_\al,  \bar{\mathcal B}_\beta\right\ra\right)_{1\leq \al,\beta\leq n}
 =\det\left( \bar\AR \right)
 \det\left( \bar{\mathcal B} \right)
 \el
 \ee
where $\bar A_1$ and $\bar B_1$ are deduced from $A_1$ and $B_1$ 
 (as given in (\ref{AB}) and (\ref{AB'})), according to whether $0\leq \tau\leq \rho$ or $\rho<\tau$, as follows:  (set $n=n_1=n_2$)
$$\bl\bar A_1&=\mbox{column of the last $n $-entries of } A_1,~~\mbox{and setting $\tau_1 =\tau $}
\\
\bar B_1&=\mbox{column of the last $n $-entries of } B_1,~~\mbox{and setting $\tau_1=\tau $}
\el $$
with
\be\label{ABscrip}\bar\AR 
 :=\left(\begin{array}{ccc}
 \bar A_1^\top (x_1) \\ \vdots \\ \bar A_{1}^\top (x_{n })\\  
 \end{array}\right)
~\mbox{and}~ \bar{\mathcal B}  
 =\left(\bar B_1(x_1),\dots,\bar B_1(x_{n }) ,
   \right).
 \ee

\end{proposition}

\proof For any $0\leq \tau$, one checks in both cases that
$$\bl
\widetilde{\mathbb L}^{\mbox{\tiny dTac}}   ( \tau,  x_i   ;\tau,  x_j)&   =\bar A^\top_1(x_i) \bar B_1(x_j).
\el$$
The rest proceeds as in the proof of Proposition \ref{Prop3.2}.\qed
 %


\section{Proof of the main Theorem \ref{Th1.2}}
 
 \noindent {\em Proof of the  joint probability formula (\ref{Prob'})}: 
 At first, notice the $n_2$ zeroes in the column $A_1$ in (\ref{AB}) and (\ref{AB'}), and so the matrix $\AR$ in (\ref{ABscrip}) contains a left-upper $0$-block of size $n_1\times n_2$. Therefore in both cases, $\rho<\tau_1<\tau_2$ and $0\leq \tau_1<\tau_2\leq \rho$, its determinant can be written as a product of determinants of two matrices ${\mathcal A}'$ of size $n_1$ and ${\mathcal A}''$ of size $n_2$:  
 \be \label{detA'}\bl  \det &(\AR)=\det\left(\begin{array}{cc}O_{n_1,n_2}&\AR'
 \\
 \AR''&~~\star_{n_2,n_1}\end{array}\right)=(-1)^{n_1n_2}\det(\AR'(\vec x))\det (\AR''(\vec y)).
 \el\ee
  Similarly, in view of the matrix $B_2$ in (\ref{AB}) and (\ref{AB'}), the matrix ${\mathcal B}$ has an upper-right $0$-block of size $n_1\times n_2$ and again $\det  ({\mathcal B})$ can be written as a product of the determinant of two matrices, a left-upper matrix ${\mathcal B}'$ of size $n_1$ and a lower-right one ${\mathcal B}''$ of size $n_2$:  
 \be\label{detB'}\bl  \det  ({\mathcal B})&=
 \det\left(\begin{array}{cc}{\mathcal B}'&O_{n_1,n_2}\\ ~~~\star_{n_2,n_1}& {\mathcal B}''\end{array} \right)=
 \det({\mathcal B}'(\vec x))\det ({\mathcal B}''(\vec y))
 \el \ee
 %
{\bf Case 1: $\rho<\tau_1<\tau_2$}. Here, from (\ref{detA'}), we find, using $n_i-r=\tau_i-\rho$,
 \be\bl
\det&(\AR'(\vec x))
\\ :=& \det\left(
 \frac{e^{-\tfrac{ x_i^2}2}}{\sqrt{\pi}} \widetilde H_{0}(x_i),\dots, \frac{e^{-\tfrac{ x_i^2}2}}{\sqrt{\pi}} \widetilde H_{n_1-r-1}(x_i),
~\vline~\frac{e^{-\tfrac{ x_i^2}2}}{\sqrt{\pi}} G^{(\tau_1)}_{n_1-r }(x_i),\dots,\frac{e^{-\tfrac{ x_i^2}2}}{\sqrt{\pi}} G^{(\tau_1)}_{n_1-1 }(x_i)\right)_{1\leq i \leq n_1}
\\\stackrel{*}=& 
\frac{1}{\sqrt{\pi}^{n_1}}\prod_{k=0}^{n_1-r-1}\frac{2^k}{k!}\prod_{j=1}^{n_1}e^{-\frac{x_j^2}2}\det\left((x_i^{j})_{{1\leq i\leq n_1}\atop{0\leq j\leq \tau_1-\rho-1}}~\vline~(
 \Phi_{j}(x_i+\tfrac \beta 2))_{{1\leq i\leq n_1}\atop{\tau_1-1\leq j\leq \tau_1-r  }} \right)
\\&\times \det \left(\begin{array}{ccc}\Id_{n_1-r} & O\\
 O&(\widetilde \Ga^{(r-1)}_{k,\ell})_{0\leq k,\ell\leq r-1}  
 \end{array}\right)
 \\=&
 \frac{1}{\sqrt{\pi}^{n_1}}\prod_{k=0}^{n_1-r-1}\frac{2^k}{k!}\prod_{j=1}^{n_1}e^{-\frac{x_j^2}2}\det(\widetilde \Ga^{(r-1)}_{k,\ell})_{0\leq k,\ell\leq r-1} 
 \widetilde \Dt_{n,\tau_1}^{(\tau_1>\rho )  } (x +\tfrac \beta 2).
 \el\label{A'}\ee
$\widetilde \Dt^{\tau_1>\rho}_{n_1,\tau_1}(x)$ was defined in (\ref{Deltatilde}). Equality $\stackrel{*}=$ above uses the fact that the first linear combination of $\widetilde H_i$ in $G_{n_1-r+k}$ (as in (\ref{C})) can be eliminated by column operations, leaving the linear combination of the $\Phi$'s to be written as product of two matrices. The second determinant in (\ref{detA'}) reads, after exchanging the $n_1$ first columns and the last $n_2-n_1$ columns:
 \be
 \bl
\det&(\AR''(\vec y))
\\&:= \det 
 \left(
\tfrac{e^{-\frac{y_i^2}2}}{\sqrt{\pi}} \widetilde H_{n_2-n_1 }(y_i) 
,\dots, \tfrac{e^{-\frac{y_i^2}2}}{\sqrt{\pi}} \widetilde H_{n_2- 1 }(y_i),\vline~
\tfrac{e^{-\frac{y_i^2}2}}{\sqrt{\pi}} \widetilde H_{0}(y_i),\dots 
,\tfrac{e^{-\frac{y_i^2}2}}{\sqrt{\pi}} \widetilde H_{n_2-n_1-1}(y_i) 
\right)_{1\leq i \leq n_2}
\\&= \tfrac{(-1)^{n_1 (n_2-n_1) }}{\sqrt{\pi}^{n_2}}
 \prod_{k=0}^{n_2-1}\frac{2^k}{k!}
 \prod_{j=1}^{n_2}e^{-\frac{y_j^2}2} 
 \Dt_{n_2}(y).
 \el
 \label{A''}\ee
 %
 %
 %
 Next, from (\ref{detB'}) and (\ref{herm1}), one finds
\be\bl\det ({\mathcal B}''^{\top}(\vec y))
&:=\det\left(
e^{-\frac{y_i^2}2} \widehat H_{0}(y_i) 
,\dots,
e^{-\tfrac{  y_i^2}2}  \widehat H_{n_2 -1}(y_i)
\right)_{1\leq i\leq n_2}
\\&=
(\tfrac{1}{2})^{n_2}  \left(\prod_{i=1}^{n_2} e^{-\frac{y_i^2}{2}}\right) \Dt_{n_2}(y)
,\el \label{B''}\ee
and
 $$
 \bl
\det ({\mathcal B}'^\top(\vec x))
 &:=\det \left({\sqrt{\pi}e^{ \tfrac{ x_i^2}2}} g_{n_2-n_1 }(\vec y, x_i),
\dots,
   {\sqrt{\pi}e^{ \tfrac{ x_i^2}2}} g_{n_2-1}(\vec y, x_i)
\right)_{1\leq i\leq n_1},\el
$$
where the $g_{n_2-1}(\vec y, x_i)$'s are the solution of the linear system (\ref{Heavilin}).  By Cramer's rule, its solution is given by  
$$
g_k(\vec y,x)=-\frac{{\mathbb V}_{X_k\curvearrowright Y_x}\tau_{n_2}}{\tau_{n_2}},~~~~0\leq k\leq n_2-1,
$$
where ${\mathbb V}_{X_k\curvearrowright Y_x}$ refers to replacing the column $X_k$ by the column $Y_x$ in the matrix: 
$$
\tau_{n_2}=\det(\widetilde H_\al(y_{i}))_{{1\leq i\leq n_2}\atop{0\leq \al \leq n_2-1}}=\det \left[X_0,\dots,X_{n_2-1}\right]
  $$
with
$$
X_\al=\left(\begin{array}{cc}
\widetilde H_\al(y_1)\\
\vdots
\\
\widetilde H_\al(y_{n_2})
\end{array}\right),~~~Y_x=\left(\begin{array}{cc}
{\mathbb H}^{\tau_2-\tau_1} (2(y_1-x))\\
\vdots
\\
{\mathbb H}^{\tau_2-\tau_1} (2(y_{n_2}-x))
\end{array}\right)
$$
So, we have, using $\widetilde H_k=2^k \bar H_k$ and formula (\ref{volformula'}) for the volume and using the so-called "Higher Fay Identity\footnote{This Fay identity has been used in \cite{AvM} and prior work.}" in equality $\stackrel{*}{=}$ below: 
$$\bl
\det& ({\mathcal B}'^{\top}(\vec x))
\\&=(\sqrt{\pi})^{n_1}\left(\prod_1^{n_1}e^{\frac{x_j^2}{2}}\right)\det \left(g_{n_2-n_1-1+\beta}(\vec y, x_\al)\right)_{1\leq \al,\beta\leq n_1}
\\ &= {(-\sqrt{\pi})^{n_1}} \left(\prod_1^{n_1}e^{\frac{x_j^2}{2}}\right)
\det\left(
\frac {{\mathbb V}_{X_{n_2-n_1-1+\beta}\curvearrowright Y_{x_\al}}\tau_{n_2}} {\tau_{n_2}}
 \right)_{1\leq \al,\beta\leq n_1}
\\ &\stackrel{*}{=}\frac{(-\sqrt{\pi})^{n_1}}{\tau_{n_2}}
\left(\prod_1^{n_1}e^{\frac{x_j^2}{2}}\right)
\left(
\prod_{\al=1}^{n_1}\BV_{X_{n_2-n_1-1+\al}\curvearrowright Y_{x_\al}}\right) \tau_{n_2}
\\&= 
\frac{(-\sqrt{\pi})^{n_1}}{\tau_{n_2}}
\left(\prod_1^{n_1}e^{\frac{x_j^2}{2}}\right)
\\
&~\times
\det\left(\!\!\begin{array}{cccccccc}
\widetilde H_{0}(y_1)&\dots& \widetilde H_{n_2-n_1-1}(y_{1})&
{\mathbb H}^{n_2\!-\!n_1} (2( y_{1}\!-\!x_1) )
&\dots&{\mathbb H}^{n_2\!-\!n_1} (2( y_{1}\!-\!x_{n_1}) )
\\
\vdots&&\vdots&\vdots&&\vdots
\\
\widetilde H_{0}(y_{n_2})&\dots& \widetilde H_{n_2-n_1-1}(y_{n_2})&
{\mathbb H}^{n_2\!-\!n_1} (2( y_{n_2}\!-\!x_1) )
&\dots&{\mathbb H}^{n_2\!-\!n_1} (2( y_{n_2}\!-\!x_{n_1}) )
\end{array}\!\!
\right)
\el
$$
\be\bl
\\&=(-1)^{\frac 12(n_2-n_1)(n_2+n_1-1)}(-1)^{-n_1}
\frac{ (\sqrt{\pi})^{n_1}\left(\prod_{k=0}^{n_2-n_1-1} {2^k} \right)2^{(n_2-n_1-1)n_1}
}
{\left(\prod_{k=0}^{n_2-1} \frac{2^k}{k!}\right)\Dt_{n_2}(y)}
\\& \qquad \times\left(\prod_1^{n_1}e^{\frac{x_j^2}{2}}\right)\mbox{Vol}(\CR(n_1,x;n_2,y))
\\&=(-2)^{\frac 12(n_2+n_1)(n_2-n_1-1)}\frac{(\sqrt{\pi})^{n_1} \prod_1^{n_1}e^{\frac{x_j^2}{2}}
 }{ \left(\prod_{k=0}^{n_2-1} \frac{2^k}{k!} \right)\Dt_{n_2}(y)}
\mbox{Vol}(\CR(n_1,x;n_2,y))
\el\label{B'}\ee
Referring to formula (\ref{2-kernel'}) in Proposition \ref{Prop3.2} and the two formulas (\ref{detA'}) and (\ref{detB'}), we obtain, upon multiplying (\ref{A'}),(\ref{A''}),(\ref{B'}),(\ref{B''}), the following expression:
\be\bl
  \frac{1}{ 2^{n_1+n_2}}&p(\tau_1,{\bf x};\tau_2,{\bf y}) 
\\ &=(-1)^{n_1n_2}\det (\AR')\det (\AR'')\det({\mathcal B}')
\det({\mathcal B}'')
\\=& 
(-1)^{n_1n_2}\frac{1}{\sqrt{\pi}^{n_1}}\prod_{k=0}^{n_1-r-1}\frac{2^k}{k!}\prod_{j=1}^{n_1}e^{-\frac{x_j^2}2}\det(\widetilde \Ga^{(r-1)}_{k,\ell})_{0\leq k,\ell\leq r-1} 
\widetilde \Dt_{n,\tau_1}^{(\tau_1>\rho )  } (x+\tfrac \beta 2 )
\\&\times
\tfrac{(-1)^{n_1 (n_2-n_1) }}{\sqrt{\pi}^{n_2}}
 \prod_{k=0}^{n_2-1}\frac{2^k}{k!}
 \prod_{j=1}^{n_2}e^{-\frac{y_j^2}2} 
 \Dt_{n_2}(y)
  \\& \times(-2)^{\frac 12(n_2+n_1)(n_2-n_1-1)}\frac{(\sqrt{\pi})^{n_1} \prod_1^{n_1}e^{\frac{x_j^2}{2}}
 }{ \left(\prod_{k=0}^{n_2-1} \frac{2^k}{k!} \right)\Dt_{n_2}(y)}
\mbox{Vol}(\CR(\tau_1,x;\tau_2,y))
\\&
\times(\tfrac{1}{2})^{n_2}  \left(\prod_{i=1}^{n_2} e^{-\frac{y_i^2}{2}}\right) \Dt_{n_2}(y) 
\\&=  \widetilde C_{\tau_1,\tau_2,r}\widetilde \Dt_{n_1,\tau_1}^{(\tau_1>\rho )  } (x+\tfrac \beta 2 ) 
\left(\prod_{i=1}^{n_2} \frac{e^{- y_i^2  }}{ \sqrt{ \pi} }\right) \Dt_{n_2}(y)
\mbox{Vol}(\CR(\tau_1,x;\tau_2,y))
 \el \label{2level}\ee
with
$$
  \widetilde C_{\tau_1,\tau_2,r}:=\frac{(-1)^{\frac 12 n_1(n_1-1)}}{(-1)^{\frac 12 n_2(n_2-1)}}\frac{\sqrt{2}^{ (n_2-n_1-1)(n_1+n_2) +(n_1-r-1)(n_1-r) }}{2^{n_2}\prod_{k=0}^{n_1-r-1}k!
 }  
\det(\widetilde \Ga^{(r-1)}_{k,\ell})_{0\leq k,\ell\leq r-1} .
$$
Then by multiplying formula (\ref{2level}) with $2^{n_1+n_2}$ gives 
$$
p(\tau_1,{\bf x};\tau_2,{\bf y}) =D(\tau_1,{\bf x};\tau_2,{\bf y})
\mbox{Vol}(\CR(\tau_1,{\bf x};\tau_2,{\bf y}))
$$
with $D(\tau_1,{\bf x};\tau_2,{\bf y})$ as in (\ref{D}) (with constant (\ref{C'})), and thus 
formula (\ref{Prob'}) for the joint probability for $\rho\leq\tau_1\leq \tau_2$, .

 \bigbreak

{\bf Case 2: $0\leq \tau_1\leq \tau_2\leq \rho$}. Here we have $n_1=n_2=r$ and so from (\ref{detA'}) and (\ref{detB'}), together with the expressions (\ref{AB'}),
%
%
one checks that, using the notation (\ref{Deltatilde}),
\be\bl\det (\AR)&=
\det\left(\begin{array}{cc}O_{r}&\AR'
 \\
 \AR''&~~\star_{r}\end{array}\right)
 \\&=(-1)^{r^2}\det(\AR'(\vec x))\det (\AR''(\vec y))
 \\&=(-1)^{r^2}\left(\prod_1^r e^{- \frac 12 (x_i^2+y_i^2) }\right)
 \\&~~~~\times\det(\widetilde H_0(y_i),\dots,\widetilde H_{r-1}(y_i))_{1\leq i\leq r}
 \det(\Phi_{ \tau_1-1}( x_i+\tfrac{\beta} 2),\dots,\Phi_{ \tau_1-r}( x_i+\tfrac{\beta} 2))_{1\leq i\leq r}
\\&=(-1)^{r^2}\left(\prod_{ 0}^{r-1}\frac{2^k}{k!}\right)\prod_1^r e^{- \frac 12 (x_i^2+y_i^2) }\widetilde \Dt_{r,\tau_1}^{(\tau_1\leq\rho)}(x+\tfrac \beta 2)\Dt_r(y),
 \el\label{detA}\ee
 and
 \be\bl  \det  ({\mathcal B})
  &=
 \det\left(\begin{array}{cc}{\mathcal B}'(x)&~~~\star_{r}\\O_{r} & {\mathcal B}''(y)\end{array} \right)
 \\&=
 \det({\mathcal B}'(\vec x))\det ({\mathcal B}''(\vec y))
 \\&=(\prod_{j=1}^r e^{  \frac 12 (x_j^2+y_j^2) })
 \det(g_0(\vec y,x_i),\dots,g_{r-1}(\vec y,x_i))_{1\leq i\leq r}
 \\
 &~~~~~\times\det(
 \Phi_{\rho-\tau_2-1}(-y_i),\dots, \Phi_{\rho-\tau_2-r}(-y_i))_{1\leq i\leq r}
  \det\left(   \widetilde \Ga_{k,\ell}\right)_{0\leq k,\ell \leq r-1}     
  \\& = \det\left(   \widetilde \Ga_{k,\ell}\right)_{0\leq k,\ell \leq r-1}(\prod_{j=1}^r e^{  \frac 12 (x_j^2+y_j^2) }) \widetilde\Dt_{r,\rho-\tau_2}^{(\tau_2\leq\rho)}(-{\bf y} ) \det (g_{j-1}(\vec y,x_i))_{1\leq i,j\leq r}.
\el \label{detB}\ee
The $g_k$'s are solution of the system (\ref{Heavilin'}), namely, as before,
 $$
g_k(\vec y,x)=-\frac{{\mathbb V}_{X_k\curvearrowright Y_x}\tau_{r}}{\tau_{r}},~~~~0\leq k\leq r-1,
$$
where ${\mathbb V}_{X_k\curvearrowright Y_x}$ refers to replacing the column $X_k$ by the column $Y_x$ in the matrix: 
$$
\tau_{r}=\det(\widetilde H_\al(y_{i}))_{{1\leq i\leq r}\atop{0\leq \al \leq r-1}}=\det \left[X_0,\dots,X_{r-1}\right],
  $$
with
$$
X_\al=\left(\begin{array}{cc}
\widetilde H_\al(y_1)\\
\vdots
\\
\widetilde H_\al(y_{r})
\end{array}\right),~~~Y_x=\left(\begin{array}{cc}
{\mathbb H}^{\tau_2-\tau_1} (2(y_1-x))\\
\vdots
\\
{\mathbb H}^{\tau_2-\tau_1} (2(y_{r}-x))
\end{array}\right),
$$
one finds, again using the "Higher Fay Identity", that
\be\bl\det (g_{j-1}(\vec y,x_i))_{1\leq i,j\leq r}
&=\frac{(-1)^r}{\tau_r}\det\left( {\mathbb H}^{\tau_2-\tau_1} (2(y_i-x_j)) \right)_{1\leq i,j\leq r }
\\&=\frac{(-1)^r  2^{r(\tau_2-\tau_1-1)} }{\Dt_r(y)\prod_0^{r-1}\frac{2^k}{k!}}
\det\left( {\mathbb H}^{\tau_2-\tau_1} ( y_i-x_j ) \right)_{1\leq i,j\leq r }
\\&=\frac{(-1)^r  2^{r(\tau_2-\tau_1-1)} }{\Dt_r(y)\prod_0^{r-1}\frac{2^k}{k!}}\mbox{Vol}(\CR(\tau_1,x;\tau_2,y)).
\el\label{detg}\ee
Thus, determinant (\ref{detB}), taking into account (\ref{detg}), yields 
\be\label{52}\bl   \det  ({\mathcal B})=&
 \det\left(   \widetilde \Ga_{k,\ell}\right)_{0\leq k,\ell \leq r-1}\frac{(-1)^r  2^{r(\tau_2-\tau_1-1)} }{\prod_0^{r-1}\frac{2^k}{k!}}(\prod_{j=1}^r e^{  \frac 12 (x_j^2+y_j^2) }) \frac{\widetilde\Dt_{r,\rho-\tau_2}^{(\tau_2\leq\rho)}(-{\bf y} ) }{\Dt_r({\bf y})}\\&\hspace*{6cm}\times\mbox{Vol}(\CR(\tau_1,x;\tau_2,y)).\el\ee
So, the two-level density (\ref{2-kernel'}) gives, by (\ref{detA}) and (\ref{52}), that  
\be\bl
\frac{1}{2^{2r}} p(\tau_1,{\bf x};\tau_2,{\bf y})
 &= \det (\AR ) \det({\mathcal B} )
 %
%
 \\&=\widetilde C'_{\tau_1,\tau_2,r}
 \widetilde\Dt_{r,\tau_1}^{(\tau_1\leq\rho)}(x+\tfrac \beta 2)
 \widetilde\Dt_{r,\rho-\tau_2}^{(\tau_2\leq\rho)}(-y )
 \mbox{Vol}(\CR(\tau_1,x;\tau_2,y)),
\el \ee
with
$$\widetilde C'_{\tau_1,\tau_2,r}:=
  (-1)^{ r(r-1)}2^{r(\tau_2-\tau_1-1)} \det\left(   \widetilde \Ga_{k,\ell}\right)_{0\leq k,\ell \leq r-1} .
$$
Multiplying with $2^{2r} $ gives 
$$p(\tau_1,{\bf x};\tau_2,{\bf y})=D(\tau_1,{\bf x}^{(\tau_1)};\tau_2,{\bf y}^{(\tau_2)}) \mbox{Vol}(\CR(\tau_1,x;\tau_2,y))
$$
with constant $C'_{\tau_1,\tau_2,r}$, as in (\ref{C'}), and thus formula  (\ref{Prob'}) for $0\leq \tau_1\leq \tau_2\leq \rho$. This ends the proof of Theorem \ref{Th1.2} for the joint probability.\qed


\bigbreak

 \noindent {\em Proof of the  single probability formula (\ref{Prob'})}:   From Proposition \ref{Prop4.3}, formulas (\ref{2-kernel}) and (\ref{ABscrip}) and looking at formulas (\ref{AB}) and (\ref{AB'}), it follows that for each of the ranges of $\tau$:

 \noindent {\bf Case 1.} For $\rho<\tau$, we have  
  \be\bl
\det (\bar\AR ({\bf x}))
  &=
  \prod_{k=0}^{n -r-1}\frac{2^k}{k!}\prod_{j=1}^{n }\frac{e^{-\frac{x_j^2}2}}{\sqrt{\pi}}\det(\widetilde \Ga^{(r-1)}_{k,\ell})_{0\leq k,\ell\leq r-1} 
 \widetilde \Dt_{n,\tau}^{(\tau>\rho )  } ({\bf x} +\tfrac \beta 2).
\\
\det (\bar{\mathcal B} ^{ }({\bf x}))
&=
(\tfrac{1}{2})^{n }  \left(\prod_{i=1}^{n } e^{-\frac{x_i^2}{2}}\right) \Dt_{n }({\bf x})
\el \label{ABbar}\ee
and thus, from (\ref{2-kernel}), 
$$\bl\frac{1}{2^n}p(\tau, {\bf x}) &=\det (\bar\AR ({\bf x}))\det (\bar{\mathcal B} ({\bf x})) 
\\&=\frac{1}{2^n}\prod_{k=0}^{n-r-1}\frac{2^k}{k!}\det(\widetilde \Ga^{(r-1)}_{k,\ell})_{0\leq k,\ell\leq r-1} \prod_{j=1}^{n }\frac{e^{- {x_j^2} }}{\sqrt{\pi}}
\widetilde \Dt_{n,\tau}^{(\tau>\rho )  } ({\bf x} +\tfrac \beta 2)\Dt_{n }({\bf x}) 
\\&= \frac{1}{2^n} D(\tau,{\bf x}; \tau,{\bf x}) 
 \el$$
 yielding $D(\tau,{\bf x}; \tau,{\bf x})$ as in (\ref{D}), with the constant $C_{\tau,\tau,r}$ as in (\ref{C'}).
\medbreak
\noindent{\bf Case 2.} For $0\leq \tau\leq \rho $, we have  
\be\bl
\det (\bar\AR ({\bf x}))
  &=\left(\prod_1^ne^{-\frac{x_i^2}2}\right)\det(\Phi_{ \tau -1}( x_i+\tfrac{\beta} 2),\dots,\Phi_{ \tau -r}( x_i+\tfrac{\beta} 2))_{1\leq i\leq r}
\\
\det (\bar{\mathcal B} ({\bf x}))
  &=\left(\prod_1^ne^{ \frac{x_i^2}2}\right)\det(
 \Phi_{\rho-\tau -1}(-x_i),\dots, \Phi_{\rho-\tau -r}(-x_i))_{1\leq i\leq r}
  \det\left(   \widetilde \Ga_{k,\ell}\right)_{0\leq k,\ell \leq r-1} 
  \el \ee
  and so 
  $$\bl\frac{1}{2^r}p(\tau, {\bf x}) & =\det (\bar\AR ({\bf x}))\det (\bar{\mathcal B} ({\bf x})) 
 \\&=\det\left(   \widetilde \Ga_{k,\ell}\right)_{0\leq k,\ell \leq r-1} \widetilde\Dt_{r,\tau }^{(\tau \leq\rho)}({\bf x}+\tfrac \beta 2)
 \widetilde\Dt_{r,\rho-\tau }^{(\tau \leq\rho)}(-{\bf x} ).
 \el$$
 Multiplying by $2^r$ yields the constant $C'_{\tau,\tau,r}$ in (\ref{C'}). This ends the proof of Theorem \ref{Th1.2} for the single probability.\qed

\bigbreak

\noindent {\em Proof of Corollary \ref{Cor1.3}:} The left hand side of (\ref{Gibbs}) can be written as 
$$\bl
  \BP&\left(\begin{array}{l} 
  {\bf z}^{(\tau_1+1)}\in d{\bf z}^{(\tau_1+1)} , 
\dots,  ~{\bf z}^{(\tau_2-1)}\in d{\bf z}^{(\tau_2-1)}\end{array}\Bigr| {\bf z}^{(\tau_1)}={\bf x}^{(\tau_1)},~
{\bf z}^{(\tau_2)}={\bf y}^{(\tau_2)}\right)
\\&\times \BP \left(\begin{array}{l} {\bf x}^{(\tau_1)}\in d{\bf x}^{(\tau_1)}  \mbox{   and  }  {\bf y}^{(\tau_2)}\in d{\bf y}^{(\tau_2)}\end{array}\right)
 \el $$
 with
 $$\bl
\BP&\left( 
  {\bf z}^{(\tau_1+1)}\in d{\bf z}^{(\tau_1+1)} , 
\dots,  ~{\bf z}^{(\tau_2-1)}\in d{\bf z}^{(\tau_2-1)} \Bigr| {\bf z}^{(\tau_1)}={\bf x}^{(\tau_1)},~
{\bf z}^{(\tau_2)}={\bf y}^{(\tau_2)}\right)
\\ & = \frac{d\mu_{{\bf x}\bf y}( {\bf z}^{(\tau_1+1 )}, \dots,  {\bf z}^{(\tau_2 -1)})}{\mbox{Vol}(\CR(\tau_1,{\bf x};\tau_2,{\bf y}))},
 \el $$
 ending the proof of Corollary \ref{Cor1.3}.\qed


\end{document}